\documentclass[aps,groupedaddress,showpacs]{revtex4}
\input epsf
\usepackage{float}

\begin{document}
\preprint{IPP-GEM-1}
\title{GEM -- An Energy Conserving Electromagnetic Gyrofluid Model}
\author{Bruce D. Scott}
\email[email: ]{bds@ipp.mpg.de}
\homepage[\\ URL: ]{http://www.rzg.mpg.de/~bds/}
\affiliation{Max-Planck-Institut f\"ur Plasmaphysik, 
		Euratom Association,
                D-85748 Garching, Germany}

\date{\today}

\begin{abstract}
The details of fluctuation free energy conservation in the gyrofluid
model are examined.  The polarisation equation relates ExB flow and eddy
energy to combinations of the potential and the density and
perpendicular temperature.  These determine the combinations which must
appear under derivatives in the moment equations so that not only
thermal free energy but its combination with the ExB energy is properly
conserved by parallel and perpendicular compressional effects.
The resulting system exhibits the same qualitative energy
transfer properties as corresponding Braginskii or Landau fluid models.
One clear result is that the numerical model built on these equations is
well behaved for arbitrarily large perpendicular wavenumber, allowing
exploration of two scale phenomena linking dynamics at the ion and
electron gyroradii.  When the numerical formulation is done in the
globally consistent flux tube model, the results with adiabatic
electrons are consistent with the ``Cyclone Base Case'' results of
gyrokinetic models.
\end{abstract}


\maketitle

\def\emskip{\hskip 1 em}
\def\hfb{\hfil\break}
\def\etc{{\it etc.}}
\def\visavis{{\it vis-a-vis}\ }
\def\ie{{\it i.e.}}
\def\eg{{\it e.g.}}
\def\etal{{\it et al}}
\def\ua{u.a.\ }
\def\dh{d.h.\ }
\def\zb{z.B.\ }
\def\bzw{bzw.\ }
\def\usw{usw.\ }

\def\idelta{$i$-delta}


\def\half{ {1\over 2} }
\def\third{ {1\over 3} }
\def\fourth{ {1\over 4} }
\def\tth{ {2\over 3} }
\def\twothirds{ {2\over 3} }
\def\threehalves{ {3\over 2} }
\def\fivehalves{ {5\over 2} }
\def\fivethirds{ {5\over 3} }
\def\sevenhalves{ {7\over 2} }
\def\threeh{\threehalves}
\def\eps{\epsilon}
 
\def\grapprox{\mathop{\lower.5ex \hbox{$\buildrel{\fivesy >}\over{\fivesy\sim}$}} \nolimits}
\def\lsapprox{\mathop{\lower.5ex \hbox{$\buildrel{\fivesy <}\over{\fivesy\sim}$}} \nolimits}
\def\grls{\mathop{\lower.5ex \hbox{$\buildrel{\fivesy >}\over{\fivesy <}$}} \nolimits}

\def\vec#1{{\bf #1}}
\def\tsr#1{{\secfnt #1}}
\def\avg#1{\left\langle #1 \right\rangle}
\def\abs#1{\left\vert #1 \right\vert}
\def\prf#1{\overline{#1}}

\def\max{{}_{{\rm max}}}
\def\min{{}_{{\rm min}}}

\def\minus{\mathop{\hbox{--}}\nolimits}

\def\re{\mathop{\rm Re}\nolimits}
\def\im{\mathop{\rm Im}\nolimits}
\def\sech{\mathop{\rm sech}\nolimits}
\def\diag{\mathop{\rm diag}\nolimits}
\def\Max{\mathop{\rm Max}\nolimits}
\def\Min{\mathop{\rm Min}\nolimits}
\def\nint{\mathop{\rm NINT}\nolimits}
\def\mod{\mathop{\rm mod}\nolimits}
\def\det{\mathop{\rm det}\nolimits}
\def\Tr{\mathop{\rm Tr}\nolimits}
\def\sign{\mathop{\rm sign}\nolimits}

\def\LBR{\left\lbrace}
\def\RBR{\right\rbrace}
\def\LB{\left\lbrack}
\def\RB{\right\rbrack}
\def\LP{\left (}
\def\RP{\right )}
\def\qq{\qquad\qquad}
\def\qqq{\qquad\qquad\qquad}
\def\Det#1{\left\vert\matrix{#1}\right\vert}

\def\pt{\partial}

\def\pzz#1{{\partial #1\over\partial z}}
\def\pxx#1{{\partial #1\over\partial x}}
\def\pyy#1{{\partial #1\over\partial y}}
\def\pww#1{{\partial #1\over\partial w}}
\def\pss#1{{\partial #1\over\partial s}}
\def\prr#1{{\partial #1\over\partial r}}
\def\prhrh#1{{\partial #1\over\partial \rho}}
\def\pthth#1{{\partial #1\over\partial \theta}}
\def\pchch#1{{\partial #1\over\partial \chi}}
\def\ppsps#1{{\partial #1\over\partial \psi}}
\def\pzeze#1{{\partial #1\over\partial \zeta}}
\def\pphph#1{{\partial #1\over\partial \phi}}
\def\ptt#1{{\partial #1\over\partial t}}
\def\pVV#1{{\partial #1\over\partial V}}
\def\phh#1{{\partial #1\over\partial \theta}}
\def\pvhvh#1{{\partial #1\over\partial \vartheta}}
\def\pxixi#1{{\partial #1\over\partial \xi}}
\def\dtt#1{{d #1\over dt}}
\def\dss#1{{d #1\over ds}}
\def\drr#1{{d #1\over dr}}
\def\pprr#1{{\partial^2 #1\over\partial r^2}}
\def\pprhrh#1{{\partial^2 #1\over\partial \rho^2}}
\def\ppss#1{{\partial^2 #1\over\partial s^2}}
\def\ppxx#1{{\partial^2 #1\over\partial x^2}}
\def\ppxy#1{{\partial^2 #1\over\partial x\partial y}}
\def\ppyy#1{{\partial^2 #1\over\partial y^2}}
\def\ppys#1{{\partial^2 #1\over\partial y\partial s}}
\def\ppzz#1{{\partial^2 #1\over\partial z^2}}
\def\pptt#1{{\partial^2 #1\over\partial t^2}}
\def\ppVV#1{{\partial^2 #1\over\partial V^2}}
\def\ppphph#1{{\partial^2 #1\over\partial \phi^2}}
\def\ppthth#1{{\partial^2 #1\over\partial \theta^2}}
\def\pphh#1{{\partial^2 #1\over\partial \theta^2}}
\def\ppvhvh#1{{\partial^2 #1\over\partial \vartheta^2}}
\def\ppxixi#1{{\partial^2 #1\over\partial \xi^2}}
\def\ppzeze#1{{\partial^2 #1\over\partial \zeta^2}}
\def\pphze#1{{\partial^2 #1\over\partial\theta\partial\zeta}}
\def\ppz#1{\partial #1/\partial z}
\def\ppx#1{\partial #1/\partial x}
\def\ppy#1{\partial #1/\partial y}
\def\ppw#1{\partial #1/\partial w}
\def\ppr#1{\partial #1/\partial r}
\def\pprh#1{\partial #1/\partial \rho}
\def\pps#1{\partial #1/\partial s}
\def\ppt#1{\partial #1/\partial t}
\def\ppV#1{\partial #1/\partial V}
\def\pph#1{\partial #1/\partial \theta}
\def\ppvh#1{\partial #1/\partial \vartheta}
\def\ppxi#1{\partial #1/\partial \xi}

\def\ddt#1{d #1/dt}
\def\pppz#1{\partial^2 #1/\partial z^2}
\def\pppx#1{\partial^2 #1/\partial x^2}
\def\pppy#1{\partial^2 #1/\partial y^2}
\def\pppr#1{\partial^2 #1/\partial r^2}
\def\ppprh#1{\partial^2 #1/\partial \rho^2}
\def\ppps#1{\partial^2 #1/\partial s^2}
\def\pppt#1{\partial^2 #1/\partial t^2}
\def\pppV#1{\partial^2 #1/\partial V^2}
\def\ppph#1{\partial^2 #1/\partial \theta^2}
\def\pppvh#1{\partial^2 #1/\partial \vartheta^2}
\def\pppxi#1{\partial^2 #1/\partial \xi^2}
\def\dddt#1{d^2 #1/dt^2}

\def\grad{\nabla}
\def\cross{{\bf \times}}
\def\div{\grad\cdot}
\def\divp{\grad_\perp\cdot}
\def\divpl{\grad_\parallel\cdot}
\def\curl{\grad\cross}
\def\dpl{\grad_\parallel}
\def\ddpl{\grad_\parallel^2}
\def\dpp{\grad_\perp}
\def\ddpp{\grad_\perp^2}
\def\delsq{\grad^2}
\def\delamb{ \mathchar"0274\hskip -.665em\mathchar"0275 }
\let\delam=\delamb
\def\lapl{\grad^2}
\def\lapldef{\ddpp=(\pt^2/\pt x^2)+K^2(\pt^2/\pt y^2)}

\def\pwww#1{{\partial #1\over\partial \vec w}}
\def\pwwpl#1{{\partial #1\over\partial w_\parallel}}
 
\def\pvv#1#2{{\partial #2\over\partial v_{#1}}}
\def\ppv#1#2{{\partial #2/\partial v_{#1}}}
\def\pvvv#1{{\partial #1\over\partial \vec v}}
\def\pvvp#1#2{{\partial #2\over\partial v'_{#1}}}
\def\ppvp#1#2{{\partial #2/\partial v'_{#1}}}
\def\pvvvp#1{{\partial #1\over\partial \vec v'}}
\def\pvvpl#1{{\partial #1\over\partial v_\parallel}}
 
\def\xunit{\vec{\hat x}}
\def\yunit{\vec{\hat y}}
\def\zunit{\vec{\hat z}}
\def\sunit{\vec{\hat s}}
\def\bunit{\vec{b}}
\def\eunit{\vec{\hat e}}
\def\nunit{\vec{\hat n}}
\def\dt{\Delta t}
\def\becomes{\leftarrow}
\def\from{\leftarrow}
\def\to{\rightarrow}
\def\fromto{\leftrightarrow}
\def\implies{\,\,\,\Longrightarrow\,\,\,}
\def\dotdot{\!:\!}

\def\meters{\,{\rm m}}
\def\invm{\,{\rm m}^{-3}}
\def\invsec{\,{\rm sec}^{-1}}
\def\cm{\,{\rm cm}}
\def\km{\,{\rm km}}
\def\invcc{\,{\rm cm}^{-3}}
\def\invcm{\,{\rm cm}^{-1}}
\def\invmm{\,{\rm mm}^{-1}}
\def\mm{\,{\rm mm}}
\def\Vcm{\,{\rm V/cm}}
\def\Acm{\,{\rm A/cm^2}}
\def\kA{\,{\rm kA}}
\def\MA{\,{\rm MA}}
\def\degk{\,{\rm K}}
\def\ergs{\,{\rm erg}}
\def\eV{\,{\rm eV}}
\def\keV{\,{\rm keV}}
\def\MeV{\,{\rm MeV}}
\def\GeV{\,{\rm GeV}}
\def\kG{\,{\rm kG}}
\def\tesla{\,{\rm T}}
\def\kW{\,{\rm kW}}
\def\MW{\,{\rm MW}}
\def\MWsqm{\,{\rm MW/m^2}}
\def\Wsqm{\,{\rm W/m^2}}
\def\radsec{\,{\rm rad/sec}}
\def\Hz{\,{\rm Hz}}
\def\kHz{\,{\rm kHz}}
\def\MHz{\,{\rm MHz}}
\def\mpersec{\,{\rm m}/{\rm sec}}
\def\msqsec{\,{\rm m^2}/{\rm sec}}
\def\cmsec{\,{\rm cm}/{\rm sec}}
\def\kmsec{\,{\rm km}/{\rm sec}}
\def\ccpersec{\,{\rm cm}^3/{\rm sec}}
\def\minutes{\,{\rm min}}
\def\yr{\,{\rm yr}}
\def\hr{\,{\rm hr}}
\def\Bar{\,{\rm bar}}
\def\sec{\,{\rm sec}}
\def\msec{\,{\rm msec}}
\def\usec{\,\mu{\rm sec}}
 
\def\bdel{\vec b\cdot\grad}
\def\Bdel{\vec B\cdot\grad}
\def\Jdel{\vec J\cdot\grad}
\def\bdot{\vec B\cdot}
\def\Bdot{\vec B\cdot}
\def\exb{\vec E\cross\vec B}
\def\jxb{\vec J\cross\vec B}
\def\uxb{\vec u\cross\vec B}
\def\vxb{\vec v\cross\vec B}
\def\wxb{\vec w\cross\vec B}
\def\ucxb{{\vec u\over c}\cross\vec B}
\def\vcxb{{\vec v\over c}\cross\vec B}
\def\wcxb{{\vec w\over c}\cross\vec B}
\def\jcxb{{\vec J\cross\vec B\over c}}

\def\vexb{\vec v_E}
\def\vpol{\vec v_p}
\def\upol{\vec u_p}
\def\vstar{\vec v_*}
\def\ustar{\vec u_*}
\def\Jstar{\vec J_*}
\def\Jpol{\vec J_p}
\def\vgradb{\vec v_{\grad B}}
\def\qstar{\vec q_\wedge}
\def\qestar{\vec q_e{}_\wedge}
\def\qistar{\vec q_i{}_\wedge}
\def\pistar{\vec\Pi_*}
\def\vR{\vec v_R}
\def\vdl{\vec v\cdot\grad}
\def\vdel{\vec v\cdot\grad}
\def\vedl{\vexb\cdot\grad}
\def\udl{\vec u\cdot\grad}
\def\udel{\vec u\cdot\grad}
\def\uidl{\vec u_i\cdot\grad}
\def\uidel{\vec u_i\cdot\grad}
\def\wdel{\vec w\cdot\grad}
\def\dedt#1{d_E #1/dt}
\def\dett#1{{d_E #1\over dt}}
\def\jpp{J_\perp}
\def\jperp{\vec\jpp}
\def\qpp{q_\perp}
\def\qperp{\vec\qpp}
\def\upp{u_\perp}
\def\uperp{\vec\upp}
\def\wpl{w_\parallel}
\def\wpp{w_\perp}
\def\wperp{\vec\wpp}
\def\vpp{v_\perp}
\def\vperp{\vec\vpp}
\def\lnb{\log B}
 
\def\rms{_{rms}}
 
\def\Jpl{J_\parallel}
\def\jpl{J_\parallel}
\def\Jpp{J_\perp}
\def\jpp{J_\perp}
\def\Jperp{\vec\Jpp}
\def\Bperp{\vec B_\perp}
\def\Apl{A_\parallel}
\def\apl{A_\parallel}
\def\App{A_\perp}
\def\app{A_\perp}
\def\Aperp{\vec\App}
\def\Epl{E_\parallel}
\def\epl{E_\parallel}
\def\Epp{E_\perp}
\def\epp{E_\perp}
\def\Eperp{\vec\Epp}
\def\upl{u_\parallel}
\def\vpl{v_\parallel}
\def\Upl{U_\parallel}
\def\vor{\grad_\perp^2\phi}
\def\kpl{k_\parallel}
\def\kkpl{k_\parallel^2}
\def\kpp{k_\perp}
\def\kperp{\vec\kpp}
\def\kkpp{k_\perp^2}
\def\xpl{{x_\parallel}}
\def\xpp{x_\perp}
\def\DD{\Delta_D}
\def\Dpl{D_\parallel}
\def\Dpp{\Delta_\perp}
\def\Depl{D_e{}_\parallel}
\def\Dipl{D_i{}_\parallel}
\def\Rpl{R_\parallel}
\def\qpl{q_\parallel}
\def\qepl{q_e{}_\parallel}
\def\qipl{q_i{}_\parallel}
\def\mupl{\mu_\parallel}
\def\mupp{\mu_\perp}
\def\nuei{\nu_{ei}}
\def\nuee{\nu_{ee}}
\def\nuii{\nu_{ii}}
\def\wpe{\omega_{pe}}
\def\wpi{\omega_{pi}}
\def\nudamp{\nu_d}
\def\zeff{Z_{\!e\!f\!f}}
\def\lmfp{\lambda_{\!m\!f\!p}}
\def\ws{{\omega_*}}
\def\wsi{{\omega_{*i}}}
\def\wn{\omega_n}
\def\wt{\omega_t}
\def\wi{\omega_i}
\def\wT{\omega_T}
\def\wp{\omega_p}
\def\wc{{\omega_c}}
\def\kappacv{{\cal K}}
\def\wcv{{\omega_B}}
\def\etai{\eta_i}
\def\taui{\tau_i}
\def\rs{\rho_s}
\def\ld{\lambda_D}
\def\Lpl{L_\parallel}
\def\Lpp{L_\perp}
\def\lcorpl{\lambda_\parallel}
\def\lcorpp{\lambda_\perp}
\def\rch{\rho_{ch}}
\def\npl{\eta_\parallel}
\def\etapl{\eta_\parallel}
\def\ald{a_L}
\def\alde{a_{Le}}
\def\aldi{a_{Li}}
\def\npp{\eta_\perp}
\def\etapp{\eta_\perp}
\def\kappapl{\kappa_\parallel}
\def\dprime{\Delta'}
\def\sk{{}_{\vec k}}
\def\sky{{}_{k_y}}
\def\gk{\gamma_k}
\def\vk{\vfl_k}
\def\nk{\nfl_k}
\def\tk{\tfl_k}
\def\dk{\Delta k}
\def\gd{\gamma_0}
\def\mwn{\Delta_n}
\def\mwh{\Delta_h}
\def\gamT{\Gamma_T}
\def\gamn{\Gamma_n}
\def\gamt{\Gamma_t}
\def\gami{\Gamma_i}
\def\gamc{\Gamma_c}
\def\gamk{\Gamma_k}
\def\gams{\Gamma_s}
\def\gaml{\Gamma_l}
\def\gamr{\Gamma_r}
 
\def\ptb{\widetilde}
\def\psifl{\widetilde\psi}
\def\phifl{\widetilde\phi}
\def\ffl{\widetilde f}
\def\fe{f_e}
\def\fefl{\widetilde f_e}
\def\fifl{\widetilde f_i}
\def\nfl{\widetilde n}
\def\hfl{\widetilde h}
\def\tfl{\widetilde T}
\def\nefl{\widetilde n_e}
\def\nifl{\widetilde n_i}
\def\tefl{\widetilde T_e}
\def\tifl{\widetilde T_i}
\def\pfl{\widetilde p}
\def\pefl{\widetilde p_e}
\def\pifl{\widetilde p_i}
\def\hefl{\widetilde h_e}
\def\vx{\widetilde v_x}
\def\vfl{\widetilde v}
\def\vefl{\widetilde \vexb}
\def\vxfl{\widetilde v_x}
\def\vyfl{\widetilde v_y}
\def\vrfl{\widetilde v_r}
\def\vppfl{\widetilde v_\perp}
\def\vplfl{\widetilde \vpl}
\def\Bfl{\widetilde \vec B}
\def\Bflpp{\widetilde B_\perp}
\def\Aplfl{\widetilde A_\parallel}
\def\Appfl{\widetilde A_\perp}
\def\Aperpfl{\widetilde {\vec A}_\perp}
\def\ufl{\widetilde u_\parallel}
\def\vorfl{\grad_\perp^2\phifl}
\def\jfl{\widetilde J_\parallel}
\def\qfl{\widetilde q_\parallel}
\def\qefl{\widetilde q_e{}_\parallel}
\def\qifl{\widetilde q_i{}_\parallel}
\def\jppfl{\widetilde J_\perp}
\def\jperpfl{\widetilde {\vec J}_\perp}
\def\Afl{\ptb A_\parallel}
\def\Jfl{\ptb J_\parallel}
\def\efl{\widetilde E_\parallel}
\def\Efl{\widetilde E_\parallel}
\def\Eppfl{\widetilde E_\perp}
\def\Eperpfl{\widetilde {\vec E}_\perp}
\def\etafl{\widetilde\eta}
\def\isatfl{\widetilde I_{{\rm sat}}}
\def\phiflfl{\widetilde\phi_{{\rm fl}}}
 
\def\teplfl{\widetilde T_e{}_\parallel}
\def\teppfl{\widetilde T_e{}_\perp}
\def\qeplfl{\widetilde q_e{}_\parallel}
\def\qeppfl{\widetilde q_e{}_\perp}
\def\tiplfl{\widetilde T_i{}_\parallel}
\def\tippfl{\widetilde T_i{}_\perp}
\def\qiplfl{\widetilde q_i{}_\parallel}
\def\qippfl{\widetilde q_i{}_\perp}

\def\tepl{ T_e{}_\parallel}
\def\tepp{ T_e{}_\perp}
\def\qepl{ q_e{}_\parallel}
\def\qepp{ q_e{}_\perp}
\def\tipl{ T_i{}_\parallel}
\def\tipp{ T_i{}_\perp}
\def\qipl{ q_i{}_\parallel}
\def\qipp{ q_i{}_\perp}

\def\peplfl{\widetilde p_e{}_\parallel}
\def\peppfl{\widetilde p_e{}_\perp}
\def\piplfl{\widetilde p_i{}_\parallel}
\def\pippfl{\widetilde p_i{}_\perp}

\def\pepl{ p_e{}_\parallel}
\def\pepp{ p_e{}_\perp}
\def\pipl{ p_i{}_\parallel}
\def\pipp{ p_i{}_\perp}


\def\phinn{ {e\phifl\over T} }
\def\nnn{ {\nfl\over n} }
\def\tnn{ {\tfl\over T} }
\def\unn{ {\ufl\over c_s} }
\def\vornn{ \rho_s^2\ddpp\phinn }
\def\jnn{ {\jfl\over ne} }
\def\qnn{ {\qfl\over nT} }
\def\psinn{ {\psifl\over B\rho_s} }

\def\ahat{\hat\alpha}
\def\ehat{\hat\eta}
\def\khat{\hat\kappa}
\def\shat{\hat s}
\def\bhat{\hat\beta}
\def\muhat{\hat\mu}
\def\epss{\hat\epsilon}
\def\bigpoint#1{
    \par\bigskip
    {\baselineskip=\normalbaselineskip
    \parindent=0 pt
    {\hfill\vbox{ #1  }\hfill}}
    \par\bigskip
    }
 
\def\jfm#1{{\it J. Fluid. Mech.} {\secfnt #1}}
\def\prl#1{{\it Phys. Rev. Lett.} {\secfnt #1}}
\def\physletta#1{{\it Phys. Lett. A} {\secfnt #1}}
\def\physlettb#1{{\it Phys. Lett. B} {\secfnt #1}}
\def\pf#1{{\it Phys. Fluids} {\secfnt #1}}
\def\pfa#1{{\it Phys. Fluids A} {\secfnt #1}}
\def\pfb#1{{\it Phys. Fluids B} {\secfnt #1}}
\def\physp#1{{\it Phys. Plasmas} {\secfnt #1}}
\def\nf#1{{\it Nucl. Fusion} {\secfnt #1}}
\def\njp#1{{\it New J. Phys.} {\secfnt #1}}
\def\cpp#1{{\it Contrib. Plasma Phys.} {\secfnt #1}}
\def\ppcf#1{{\it Plasma Phys. Contr. Fusion} {\secfnt #1}}
\def\plasphys#1{{\it Plasma Phys.} {\secfnt #1}}
\def\revpp#1{{\it Rev. Plasma Phys.} {\secfnt #1}}
\def\iaea#1#2{in {\it Plasma Physics and Controlled Nuclear Fusion
    Research #1}, Proceedings of the #2th International Conference}
\def\EPS#1#2#3{in {\it Proceedings of the
{#1}th European Conference on Controlled Fusion and Plasma Physics,
{#2}, {#3}} (European Physical Society, {#2}, {#3})}
\def\jcp#1{{\it J. Comput. Phys.} {\secfnt #1}}
\def\jetp#1{{\it Sov. Phys. JETP} {\secfnt #1}}
\def\sovjpp#1{{\it Sov. J. Plasma Phys.} {\secfnt #1}}
\def\jnm#1{{\it J. Nucl. Mat.} {\secfnt #1}}
\def\rsi#1{{\it Rev. Sci. Inst.} {\secfnt #1}}
\def\adv#1{{\it Adv. Phys.} {\secfnt #1}}
\def\apjl#1{{\it Astrophys. J. Lett.} {\secfnt #1}}
\def\apj#1{{\it Astrophys. J.} {\secfnt #1}}
\def\aa#1{{\it Astron. Astrophys.} {\secfnt #1}}
\def\vol#1{\ {\secfnt #1}}

\def\skp{_{\kperp}}
\def\ppb#1{\pt #1/\pt b}
\def\pbb#1{{\pt #1\over\pt b}}
\def\phigfl{\phifl_G}
\def\vorgfl{\Omegafl_G}
\def\vorggfl{\ptb{\widehat\Omega}_G}
\def\ddppg{\half\widehat\ddpp}
\def\ddppgg{\widehat{\widehat\grad}{}^2_\perp}
\def\bperp{\vec b_\perp}
\def\bbdl{\bperp\cdot\grad}
\def\pxxmu#1{{\pt #1\over\pt x^\mu}}
\def\pxxnu#1{{\pt #1\over\pt x^\nu}}
\def\aldz{a_{Lz}}
\def\aldi{a_{Li}}
\def\alde{a_{Le}}
\def\ald{a_{L0}}
\def\nucv{\nu_{{\cal K}}}

\def\bb{\vec B}
\def\pyyk#1{{\pt #1\over\pt y_k}}
\def\ppyyk#1{{\pt^2 #1\over\pt y_k^2}}
\def\vex{\ptb v_E^x}
\def\bbx{\ptb b^x}
\def\vey{\ptb v_E^y}
\def\bby{\ptb b^y}
\def\ffl{\ptb f}
\def\gfl{\ptb g}
\def\Bfl{\ptb B}
\def\Omegafl{\ptb\Omega}
\def\shatfl{\ptb S}
\def\fexp{\eta}

\def\nefl{\widetilde n_e}
\def\teplfl{\widetilde T_e{}_\parallel}
\def\teppfl{\widetilde T_e{}_\perp}
\def\qeplfl{\widetilde q_e{}_\parallel}
\def\qeppfl{\widetilde q_e{}_\perp}
\def\nifl{\widetilde n_i}
\def\Nfl{\widetilde N}
\def\tiplfl{\widetilde T_i{}_\parallel}
\def\tippfl{\widetilde T_i{}_\perp}
\def\qiplfl{\widetilde q_i{}_\parallel}
\def\qippfl{\widetilde q_i{}_\perp}
\def\nzfl{\widetilde n_z}
\def\tzfl{\widetilde T_z}
\def\uzfl{\widetilde u_z{}_\parallel}
\def\qzfl{\widetilde q_z{}_\parallel}
\def\tzplfl{\widetilde T_z{}_\parallel}
\def\tzppfl{\widetilde T_z{}_\perp}
\def\qzplfl{\widetilde q_z{}_\parallel}
\def\qzppfl{\widetilde q_z{}_\perp}

\def\tepl{ T_e{}_\parallel}
\def\tepp{ T_e{}_\perp}
\def\qepl{ q_e{}_\parallel}
\def\qepp{ q_e{}_\perp}
\def\tipl{ T_i{}_\parallel}
\def\tipp{ T_i{}_\perp}
\def\qipl{ q_i{}_\parallel}
\def\qipp{ q_i{}_\perp}
\def\tzpl{ T_z{}_\parallel}
\def\tzpp{ T_z{}_\perp}
\def\qzpl{ q_z{}_\parallel}
\def\qzpp{ q_z{}_\perp}

\def\peplfl{\widetilde p_e{}_\parallel}
\def\peppfl{\widetilde p_e{}_\perp}
\def\piplfl{\widetilde p_i{}_\parallel}
\def\pippfl{\widetilde p_i{}_\perp}
\def\pzplfl{\widetilde p_z{}_\parallel}
\def\pzppfl{\widetilde p_z{}_\perp}

\def\pepl{ p_e{}_\parallel}
\def\pepp{ p_e{}_\perp}
\def\pipl{ p_i{}_\parallel}
\def\pipp{ p_i{}_\perp}
\def\pzpl{ p_z{}_\parallel}
\def\pzpp{ p_z{}_\perp}

\def\Nzfl{\LB n_z+\nzfl\RB}
\def\Tzplfl{\LB T_z+\tzplfl\RB}
\def\Tzppfl{\LB T_z+\tzppfl\RB}
\def\Pzplfl{\LB p_z+\pzplfl\RB}
\def\Pzppfl{\LB p_z+\pzppfl\RB}

\def\chiv{\hat\chi}

\def\kkpp{k_\perp^2}

\def\uexb{\vec u_E}
\def\wexb{\vec w_E}
\def\Wexb{\vec W_E}

\def\uedl{\uexb\cdot\grad}
\def\wedl{\wexb\cdot\grad}
\def\Wedl{\Wexb\cdot\grad}

\def\fhat{\vec{\widehat F}}
\def\fdel{\fhat\cdot\grad}

\def\nupp{\nu_\perp}
\def\nupl{\nu_\parallel}
\def\nuzpl{\eta_z{}_\parallel}
\def\nuepl{\eta_e{}_\parallel}

\section{Introduction -- Gyrofluid Energy Conservation in General
\label{sec:intro}}

Gyrofluid models, whose most prominent
application has been to tokamak core
turbulence as exemplified by the Cyclone Base Case \cite{dimits},
were originally constructed to incorporate finite ion
gyroradius effects at arbitrary order into simple computations of
turbulence occurring in largely two dimensional fluid experiments
\cite{knorr}.  To treat ion temperature gradient (ITG) turbulence the
temperatures were incorporated and the model acquired several new
advection terms, producing nonlinearities as well as drift frequency
corrections, resulting from the effect of temperature fluctuations on
the gyroaveraging operator \cite{dorland,beer}.  However, although the
nonlinearities were incorporated, much of the analysis involved linear
frequencies and growth rates and the nonlinearities were added largely
as an afterthought.  Specifically, there was no complete
energetic analysis of
the type familiar from drift wave turbulence work
\cite{wakhas,waltz,ssdw,dalfloc}.  
The GEM model was introduced previously in the context of energetic
considerations, including the correspondence between fluid drift and
gyrofluid models under drift ordering \cite{eps03}.  We develop the
energetics for the six-moment model including temperature and
parallel heat flux dynamics herein, placing this model on a secure
energetic footing for the first time.

Under drift ordering \cite{driftordering}, this energetics involves
``fluctuation free energy,'' in which  the thermal free energy enters as
the average squared amplitude of the density fluctuations \cite{wakhas},
with additional contributions from the average squared amplitude of the
temperature fluctuations in the appropriately generalised models
\cite{ssdw}.  The rest of the free energy is made up of contributions
due to the ExB energy involving the electrostatic potential, the
magnetic energy involving the parallel magnetic potential, and the
parallel free energy in not only the parallel velocities but also the
parallel heat fluxes \cite{dalfloc}.
A properly constructed model should conserve this free
energy in all processes except those involving clearly identifiable
sources (gradients) and sinks (dissipative processes such as
resistivity, thermal conduction, or Landau damping).  Many models
neglect this consideration
because small errors in the energetics lead simply to
negligible contributions to the growth rate in a linear model.  However,
if the model is to be useful in a turbulent setting, the energetics must
be consistent in order to achieve a reliable saturated state in which
the salient energetic processes and the turbulent transport can be
statistically measured.

The processes linking the thermal free energy to the ExB turbulence are
of physical importance because the free energy sources and sinks are not
in the perpendicular equation of motion.
The energy source given by the general
profile gradient, is in the equations for the thermal state variables
(density, temperatures).  The dissipation processes are in the equations
for 
the parallel flux variables (current, heat fluxes).  
In saturation,
the ExB energy itself is maintained as a statistical balance between
various conservative transfer effects which connect to these sources and
sinks in the other parts of the dynamics (this neglects
certain rotation damping processes which are often
not considered).  In this balance it is important that the
conservative nature of such processes such as shear Alfv\'en dynamics or
interchange effects is maintained in the model.  The simplest example is
an isothermal two dimensional magnetohydrodynamical (MHD) interchange
model, given by 
\begin{equation}
{n_iM_i c^2\over B^2}\ptt{}\ddpp\phifl = - T_e\kappacv(\nefl)
\label{eqmhdvor}
\end{equation}
\begin{equation}
\ptt{\nefl} + \vedl \LP n_e+\nefl\RP = n_e\kappacv(\phifl)
\label{eqmhdne}
\end{equation}
where the perpendicular Laplacian and curvature operator are defined by
\begin{equation}
\ddpp = -\div[\bunit\cross(\bunit\cross\grad)]
\end{equation}
\begin{equation}
\kappacv = \div\LB {c\over B^2}(\bb\cross\grad)\RB
\end{equation}
respectively,
with $\bb=B\bunit$ the equilibrium magnetic field.  
The ExB velocity is given by
\begin{equation}
\vexb = {c\over B}\bunit\cross\grad\phifl
\label{eqvexb}
\end{equation}
The curvature terms in Eqs.\ (\ref{eqmhdvor},\ref{eqmhdne}) respectively
represent quasistatic compression of the diamagnetic current and the ExB
velocity. 

Under drift ordering \cite{driftordering}, the background parameters are
constants except where operated upon by $\vedl$, the ExB advection.
The
magnetic field is treated as constant except for the existence of
$\kappacv()$. 
In the advection term,
the velocity $\vexb$ is treated as divergence free; the finite ExB
divergence is accounted for by $\kappacv(\phifl)$.
If we multiply these
equations by $-\phifl$ and $(T_e/n_e)\nefl$, respectively, and integrate
over the entire spatial domain, we find, neglecting surface terms,
\begin{equation}
\ptt{}\int d\Lambda\,{n_iM_i\over 2}{c^2\over B^2}\abs{\dpp\phifl}^2
	= - \int d\Lambda\,T_e\nefl\kappacv\LP\phifl\RP
\label{eqexbenergy}
\end{equation}
\begin{equation}
\ptt{}\int d\Lambda\,{n_eT_e\over 2} \LP{\nefl\over n_e}\RP^2
	= - \int d\Lambda\,T_e\nefl\vedl \log n_e
	+ \int d\Lambda\,T_e\nefl\kappacv\LP\phifl\RP
\label{eqthermalenergy}
\end{equation}
where $\int d\Lambda$ by itself gives the total volume.
These two lines give the evolution of the ExB drift and thermal free
energy, respectively.  The interchange effect represented by
$\kappacv()$ transfers free energy between these two pieces
conservatively, as we would expect from a compressional process, because
$\kappacv()$ is at once a total divergence and a first order linear
differential operator.  The source is given by the advection of the
background gradient, proportional to the average flux.  This model will
saturate only if there is a loss process at the boundary, or some
nonlinear effect enters to cause the source to go to zero, or some
additional effect is considered in the model by which an explicit sink
term appears to balance the source (a detailed analysis of how this
functions in a three dimensional drift Alfv\'en model is given in Ref.\
\cite{dalfloc}).

In a gyrofluid model, there is no equation for the vorticity explicitly
involving time derivatives. Instead, at the same level of sophistication
as above, there is an evolution equation for
each of the species' gyrocenter densities, as opposed to space
densities.  
In each density equation, the potential is gyroaveraged
using a suitable convolution operator which is described by a kernel in
Fourier space:
\begin{equation}
\phigfl=G\LP\phifl\RP 
	= \sum_{\kperp} G\skp \phifl\skp\,e^{i\kperp\cdot\vec x}
\label{eqgyroaveraging}
\end{equation}
The usual form for $G\skp$ is $\Gamma_0^{1/2}(b_i)$, which is an average
of the single particle form $J_0(\kpp\vpp/\Omega_i)$ averaged over the
perturbed distribution function \cite{dorland}.  
The argument is $b_i=\kkpp\rho_i^2$, where
$\rho_i$ is the thermal gyroradius and $\Omega_i$ is the gyrofrequency.
Unless phenomena below the ion scale $\rho_i$ are considered, electron
gyroradius effects are usually ignored, so that $G\skp=1$ for them.
The potential is given by a polarisation equation which
equates the two space densities, each given by a combination of the
gyrocenter and polarisation densities, the latter involving the
gyroscreened potential.  The equations appear as \cite{knorr}:
\begin{equation}
\ptt{\nefl} + \vedl\LP n_e+\nefl\RP
= n_e \kappacv\LP\phifl\RP - {T_e\over e}\kappacv\LP\nefl\RP
\label{eqne}
\end{equation}
\begin{equation}
\ptt{\nifl} + \uedl\LP n_i+\nifl\RP
= n_i \kappacv\LP\phigfl\RP + {T_i\over e}\kappacv\LP\nifl\RP
\label{eqni}
\end{equation}
\begin{equation}
{G\LP\nifl\RP\over n_i} + {e\over T_i} \rho_i^2\ddpp\phifl
	= {\nefl\over n_e}
\label{eqpolarisation}
\end{equation}
Gyroscreening is distinct from gyroaveraging; in general this is given
by another operator whose usual kernel in $\kperp$-space is
$\Gamma_0(b_i)-1$.  The low-$\kpp$ form of this is just $-b_i$, which in
real space yields the $\rho_i^2\ddpp$ form used in Eq.\
\ref{eqpolarisation}.  

It is important to note however that
the same operator is involved in the gyroaveraging of the potential
(Eq.\ \ref{eqgyroaveraging}) as in the conversion of the gyrocenter
density, $\nifl$, to the ion space density given by the left side of
Eq.\ (\ref{eqpolarisation}).  For the ions, the operator is $G$; for the
electrons, it is unity.  Consistent with this,
the ExB advection of the ion density occurs with the
gyroaveraged potential, with velocity
\begin{equation}
\uexb = {c\over B}\bunit\cross\grad\phigfl
\label{equexb}
\end{equation}
while the electrons are advected by
the ``bare'' version, the same $\vexb$ as in Eq.\ (\ref{eqvexb}).
The curvature operator acts on the total force potential of each
species, respectively, i.e., the diamagnetic velocity divergences are
necessarily kept in the model, which cannot take the MHD form if both
pressures are to contribute to charge separation.

The point to be made here is the way in which the polarisation equation,
Eq.\ (\ref{eqpolarisation}), determines the acceptable form for many of
the differential operators in the moment equations.  The ExB energy and
the electron and ion thermal free energies are given by
\begin{equation}
U_E = \int d\Lambda\,{n_iM_i\over 2}{c^2\over B^2}\abs{\dpp\phifl}^2
\qquad
U_e = \int d\Lambda\,{n_eT_e\over 2}\LP{\nefl\over n_e}\RP^2
\qquad
U_i = \int d\Lambda\,{n_iT_i\over 2}\LP{\nifl\over n_i}\RP^2
\label{eqenergies}
\end{equation}
respectively.
Using Eq.\ (\ref{eqpolarisation}) and the Hermitian property of $G$, we
may recast $U_E$ as 
\begin{equation}
U_E = e{\phigfl \nifl\over 2} - e{\phifl\nefl\over 2}
\end{equation}
which shows the ion and electron contributions separately.
By similar means, the time derivatives are given by
\begin{equation}
\ptt{U_E} = e\phigfl\ptt{\nifl} - e\phifl\ptt{\nefl}
	= - \int d\Lambda\,T_e\nefl\kappacv\LP\phifl\RP
	- \int d\Lambda\,T_i\nifl\kappacv\LP\phigfl\RP
\end{equation}
\begin{equation}
\ptt{U_e} 
	= - \int d\Lambda\,T_e\nefl\vedl \log n_e
	+ \int d\Lambda\,T_e\nefl\kappacv\LP\phifl\RP
\end{equation}
\begin{equation}
\ptt{U_i} 
	= - \int d\Lambda\,T_i\nifl\uedl \log n_i
	+ \int d\Lambda\,T_i\nifl\kappacv\LP\phigfl\RP
\end{equation}
From this we can see that, since both polarisation and thermal energy
for each species now follow from its density equation,
the density and corresponding gyroaveraged potential must appear
together under spatial derivatives in all conservative processes
in order for the energetics to remain consistent.  
For the ions this is $\phigfl+(T_i/n_i e)\nifl$, while for the electrons it
is $\phifl-(T_e/n_e e)\nefl$.  Thus, for the ions, the combination
$\phigfl+(T_i/n_i e)\nifl$ multiplied by the curvature term in Eq.\
(\ref{eqni}) yields a term which vanishes under the integration,
conserving the free energy, and the same occurs for the electrons upon
multiplying the curvature term in Eq.\ (\ref{eqne}) by 
$\phifl-(T_e/n_e e)\nefl$.

This is relatively trivial for such a one-moment gyrofluid model,
but the central point is clear: the gyrofluid closure arising from
gyroaveraging must be done the same way in the polarisation and in the
moment equations.  That is, in this case, the operator $G$ in the
gyroaveraged potential for a given species must be the same as the one
operating upon the corresponding density in the polarisation equation.
In a model which incorporates the temperatures, this extends to another
gyroaveraging operator acting upon the temperature, and the same one
acting upon the potential producing extra finite gyroradius advection
terms, plus corrections to the temperature wherever the latter appears
under parallel gradients or curvature terms.  This is much more involved
and will be treated in the next two sections, one discussing the
problems with the currently standard gyrofluid model, and the next one
formulating the GEM model.  Then, the last section shows the
applications of GEM to the damping of kinetic shear Alfv\'en waves, to
the hyperfine electron gyroradius scale turbulence problem, and to the
Cyclone Base Case which provides the standard benchmark.

\section{Drift Ordering and Normalisation Conventions
\label{sec:norm}}

With many constant factors containing the physical units to be carried
about while manipulating the equations, it is convenient to go to a
system of normalised units.  While this is somewhat arbitrary, two
examples are useful in elucidating the salient units for gyrofluid
turbulence.  One of these is the prospect of force balance in the
electrons along the magnetic field.  There are several effects which
affect the evolution of the electric current in the electron Ohm's law,
but these all act to mediate the response of the electrons to the two
static parallel forces: the pressure gradient and the static part of the
parallel electric field, given by 
\begin{equation}
n_e e\dpl\phi - \dpl p_e 
= p_e \LP{e\dpl\phi\over T_e} - \dpl\log p_e\RP
\label{eqadiabaticforce}
\end{equation}
assuming small disturbances from the equilibrium, and a field line
geometry in which the parallel gradient for finite sized
disturbances is not allowed to vanish (see below),
a quasistatic
balance between these two forces yields the following relationship
between the disturbances:
\begin{equation}
{e\phifl\over T_e} = {\pefl\over p_e}
\end{equation}
If the parallel responses also equalises the temperature along the field
lines, we then have the combination
\begin{equation}
{e\phifl\over T_e} = {\nefl\over n_e}\qquad\qquad
{\tefl\over T_e} = 0
\label{eqadiabaticelectrons}
\end{equation}
This situation is called adiabatic electrons, and the response to the
force in Eq.\ (\ref{eqadiabaticforce}) is called the adiabatic
response.  The importance of the pressure force is what departs gradient
driven turbulence in general from the world of MHD.  The adiabatic
response indicates the useful units for the electrostatic potential and
all the thermodynamic state variables, which will be scaled in terms of
$T_e/e$, $n_e$, or $T_e$ following the forms in Eq.\
(\ref{eqadiabaticelectrons}). 

The other useful example is closer to the idea of the gyrofluid
formulation: the relationship between ion inertia and polarisation.  In
the low-$\kpp$ limit, the ExB energy is the same as the fluid one,
\begin{equation}
U_E = n_iM_i{v_E^2\over 2} 
= {n_iM_i\over 2}{c^2\over B^2}\abs{\dpp\phifl}^2
\end{equation}
given in Eqs.\ (\ref{eqexbenergy},\ref{eqenergies}).  The
functional derivative of 
$U_E$ with respect to $\phifl$ leads to the polarisation density 
\begin{equation}
\Omega = {n_iM_i c^2\over B^2}\ddpp\phifl
\end{equation}
given in Eq.\ (\ref{eqpolarisation}).  Expressing $\phifl$ in terms of
$T_e/e$ and the densities in terms of $n_e e$, we find
\begin{equation}
{\Omega\over n_e e} = {n_i\over n_e} \rs^2\ddpp{e\phifl\over T_e}
\end{equation}
where $\rs$ is the drift scale given by
\begin{equation}
\rs^2 = {T_e M_i c^2\over e^2 B^2}
\label{equation}
\end{equation}
This is equally well described in terms of the ion gyroradius,
\begin{equation}
{\Omega\over n_e e} = {n_i\over n_e} {T_e\over T_i}
\rho_i^2\ddpp{e\phifl\over T_e}
\end{equation}
The role of the drift scale remains, however, even if $T_i\to 0$,
because it is a purely inertial phenomenon.  With several species
present, $T_e$ and $\rs$ make good choices for the basis of the
normalisation scheme, and we will use them herein.

We work in terms of an arbitrary number of charged fluids, each with a
particular background density and temperature, and mass and charge state
per particle.  In terms of the electron density and temperature, $n_e$
and $T_e$, the mass of a main ion species $M_i$, and the unit of charge,
$e$, we have a normalised background charge density
\begin{equation}
a_z = {n_z Z\over n_e}
\end{equation}
temperature to charge ratio
\begin{equation}
\tau_z = {T_z\over Z T_e}
\end{equation}
and mass to charge ratio
\begin{equation}
\mu_z = {M_z\over Z M_i}
\end{equation}
for each species labelled by $z$.
For electrons, these are $a_e = \tau_e = -1$ and $\mu_e = - m_e/M_i$.
For each species, the background mass density is given by $a_z\mu_z$,
pressure by $a_z\tau_z$, and squared gyroradius by
$\rho_z^2=\mu_z\tau_z$, in units of $n_e M_i$, $p_e=n_eT_e$, and
$\rs^2$, respectively.  In fusion applications the main ion is often
considered to be deuterium, and so the latter mass $M_D$ is used for
$M_i$.  However, with discharge experiments run at fusion-relevant
normalised parameters using several disparate 
ion types becoming increasingly important \cite{tjk}, it is important
not to make this choice automatic.

We work under drift ordering, also called gyrokinetic
ordering.  There are two basic perpendicular length scales: the drift
scale $\rs$ and the background profile scale $\Lpp$.  Drift ordering
assumes their ratio to be small, 
\begin{equation}
\delta = \rs/\Lpp \ll 1
\end{equation}
where $\delta$ is called the drift parameter.  It also assumes that the
relative amplitude of all the disturbances is small by this same order,
for example,
\begin{equation}
{e\phifl\over T_e}\sim {\nzfl\over n_z} \sim {\tzfl\over T_z} 
=O(\delta)
\end{equation}
and similarly for the parallel flux variables (velocities, heat fluxes)
in terms of the sound speed $c_s$ given by 
\begin{equation}
c_s^2 = T_e/M_i
\end{equation}
It also takes a ``maximal'' ordering with respect to perpendicular
wavenumbers and the drift scale,
\begin{equation}
\kpp\rs\sim 1
\end{equation}
while assuming a flute/drift ordering for the parallel wavenumber,
\begin{equation}
\kpl/\kpp = O(\delta)
\end{equation}
The significance of these statements taken together is that the dynamics
can be very nonlinear, in the sense that $\vedl\sim\ppt{}$, even though
the disturbance amplitude remains small.  It therefore follows that all
nonlinearities are dropped except for the quadratic ones represented by
ExB advection ($\vedl$ in a fluid model, additionally all its finite
gyroradius relatives in a gyrofluid model) and, in an electromagnetic
model, the ``magnetic flutter'' nonlinearities represented by the
contributions due to the magnetic disturbances in the parallel gradient
$\dpl$.  With the flute ordering we assume that the disturbed and
undisturbed parallel gradient pieces are of similar size.

An important implication of drift ordering on the treatment of the
geometry concerns the divergences of the various perpendicular drift
fluxes.  The ExB 
velocity in an inhomogeneous magnetic field has a finite divergence, so
that both $\vedl n$ and $n\div\vexb$ would enter a fluid model.
However, under drift ordering, both the profile and disturbance are
advected at the same order by $\vexb$, while the factor of $n$
multiplying the divergence is treated as a constant parameter.  This
forces an explicit split between the advection and divergence terms.
Due to their various cancellations \cite{tsai,hinton}, the diamagnetic
fluxes enter only in 
the divergences, so this splitting
concerns solely the ExB velocity in a fluid
model, plus the associated finite gyroradius forms in a gyrofluid model.
Under drift ordering, then, the advection term appears as a pure Poisson
bracket form (between the two perpendicular coordinates in a field
aligned treatment, or between all three pairs in general),
multiplied by a constant coefficient, and the divergence term appears as
a curvature term (through $\kappacv$, as above), whose properties are
that (1) it is a pure divergence, (2) it is a first order differential
operator with the linearity property, and (3) all curvature terms are
linear in the dependent variables.

The further impact of drift ordering on the treatment of the magnetic
geometry is summarised in the statement that if the three coordinates
are aligned to the magnetic field such that one of them is parallel
($s$), one is radial ($x$, across flux surfaces, down the gradient), and
the third ($y$) has vanishing projections to both the equilibrium
magnetic field and the background gradient, then the variation of the
geometry is retained only in $s$.  Field aligning means that only one
contravariant component of $\bb$ is nonvanishing, in this case $B^s$.
These are the basic statements of field aligned coordinates and magnetic
flux tube geometry, as explained elsewhere \cite{beergeom,fluxtube}.
The principal consequence is that the perpendicular Laplacian $\ddpp$
involves only $x$ and $y$, and the metric coefficients depend only upon
$s$.  Note especially that this includes the variation of the field
strength $B$, whose variation along the magnetic field is retained but
commutes with $\ddpp$ (hence, $B^2$ appears as a coefficient in the
normalised gyroradius in all gyroaveraging and gyroscreening
operations).  Although the equations of the standard gyrofluid model and
of GEM are expressed in covariant terms, the flux tube geometry enters
when they are discretised in a numerical representation, and their
essential coordinate dependence is reflected in the abovementioned split
between advection by and divergence of the drift velocity.  We will save
further comment on this for the sections (below) describing the
numerical scheme.  

Thermodynamic state variables are normalised in terms of their
background quantities, the electrostatic potential in terms of $T_e/e$,
and the parallel magnetic potential in terms of $B_0\rs$.  Additionally,
a factor of $\delta$ is folded into all the normalisations, leaving the
ExB advective nonlinearities coefficientless.  This is expressed as
\begin{equation}
\phifl\from \delta^{-1}{e\phifl\over T_e}\qquad\qquad
\Afl\from \delta^{-1}{\Afl\over B_0\rs\beta_e}
\end{equation}
for the potentials, and
\begin{equation}
\nzfl\from \delta^{-1}{\nzfl\over n_z}\qquad\qquad
\tzfl\from \delta^{-1}{\tzfl\over T_z}
\end{equation}
\begin{equation}
\uzfl\from \delta^{-1}{\uzfl\over c_s}\qquad\qquad
\qzfl\from \delta^{-1}{\qzfl\over n_z T_z c_s}
\end{equation}
for the state and parallel flux variables respectively, where 
\begin{equation}
\beta_e = {4\pi p_e\over B^2}
\end{equation}
is the electron dynamical beta (this enters through Ampere's law).
The density and temperature are therefore treated on an equal footing
with respect to their flux variables, the velocity and heat flux,
respectively.  Additionally, in both GEM and the standard gyrofluid
model, parallel and perpendicular temperatures and parallel-parallel and
perp-parallel heat fluxes are treated separately, leaving a six moment
model, for both ions and electrons as previously \cite{gyroem}, and for
all species herein.

We leave $\dpl$ normalised in terms of $\Lpp$.  Hence, the size of the
contravariant magnetic unit vector component $b^s$ is comparable to
$\Lpp/qR$, 
where $q$ is the magnetic pitch parameter
(toroidal/poloidal magnetic field, contravariant component ratio), $R$
is the toroidal major radius, and $2\pi qR$ gives the 
connection length along the magnetic field lines.  
The parameters governing
core or edge turbulence result from competition between ExB turbulence
and parallel dynamics, and so the scale ratios enter the ion inertia and
curvature parameters according to
\begin{equation}
\epss=\LP{qR\over\Lpp}\RP^2 \qquad\qquad \wcv={2\Lpp\over R}
\end{equation}
respectively.
These lead to the
parameters governing the parallel electron dynamics,
\begin{equation}
\bhat=\beta_e\epss \qquad\qquad \muhat={m_e\over M_i}\,\epss
\qquad\qquad C={0.51\nu_e\over c_s/\Lpp}\,\muhat
\label{eqparms}
\end{equation}
These are the drift Alfv\'en parameter, the electron inertia parameter,
and the drift wave collisionality parameter, respectively.  For core
turbulence $C$ and $\muhat$
are small, and $\bhat$ can be in the range from small values to several
times $0.1$ in high performance tokamaks.  
For edge turbulence $C$ and $\muhat$
are larger than unity, and the turbulence becomes electromagnetic for
$\bhat$ near or larger than unity.  Ideal and resistive ballooning
regimes occur when $\bhat\wcv>1$ or $C\wcv>1$, respectively.  
The magnetic shear parameter is $\shat$, nominally given by $d\log
q/d\log r$ where $r$ is the minor radius but generalisable to arbitrary
geometry (cf.\ Ref.\ \cite{fluxtube}).  It is usually of order unity.
Under the shifted metric fluxtube geometry, $\shat$ enters only through
the shifts incurred in the $y$-coordinate while taking derivatives with
respect to $s$ \cite{shifted}.
Further
details on the significance of these parameters may be found in
Ref.\ \cite{eps03}.


\section{Energetic Problems with the Standard Gyrofluid Model
\label{sec:standard}}

{\def\epss{\null}

Most of the problems with the standard gyrofluid model
\cite{dorland,beer} arise from the finite Larmor radius (FLR) terms.
The closure approximations are done term by term in a reasonable way,
but with subtle differences in the polarisation equation and in the
gyrofluid moment variable equations.  In the polarisation equation the
gyroaveraging of the distribution function is computed from density and
temperature moments of a perturbed Maxwellian approximation.  In the
moment equations the gyroaveraging is done on the potential
(concentrating here upon $\phifl$), which is subject to derivatives and
then to the moment integrals.  These two procedures are in general
different unless one approaches them simultaneously with energy
conservation in mind.  For this reason, the result that the model has
inconsistencies is not so unreasonable, particularly considering that
its motivation starts with linear theory and only secondarily adds the
nonlinearities one needs for turbulence.  Additional to this is an
inconsistency in
treating the higher moments which arise from the curvature terms: a
perturbed Maxwellian model is used for those terms while the model
itself retains the parallel heat flux moments as dynamical variables.
These difficulties are treated in turn, with the FLR terms first.  The
standard gyrofluid model is sufficiently well constructed that these
repairs are in the end a minor matter.

There are three gyroaveraging operators used in the model, given by
Eqs.\ (20,26,27) of Ref.\ \cite{beer},
\begin{equation}
\phigfl = \Gamma_0^{1/2}\phifl\qquad
\ddppg\phigfl = b{\pt\Gamma_0^{1/2}\over\pt b}\phifl\qquad
\ddppgg\phigfl = b{\pt^2\over\pt b^2}\LP b\Gamma_0^{1/2}\RP\phifl
\end{equation}
assuming a field aligned coordinate system and a Fourier representation
in the two perpendicular coordinates, such that $b=\kkpp\rho_i^2$ (the
model concentrates on a single ion species and leaves the electrons to
be adiabatic).  We relabel these in terms of $\Gamma_1$, $\Gamma_2$, and
$\Gamma_3$, given by
\begin{equation}
\Gamma_1=\Gamma_0^{1/2}\qquad\qquad
\Gamma_2=b{\pt\Gamma_1\over\pt b}\qquad\qquad
\Gamma_3={b\over 2}{\pt^2\over\pt b^2}\LP b\Gamma_1\RP
\end{equation}
and recast the gyroaveraged and FLR corrected potentials as
\begin{equation}
\phigfl = \Gamma_1\phifl\qquad\qquad
\vorgfl = \Gamma_2\phifl\qquad\qquad
\vorggfl = \Gamma_3\phifl
\end{equation}
for clarity.  Note the factor of two inserted into the definition for
$\Gamma_3$, so that $\vorgfl$ and $\vorggfl$ both have the same
low-$\kpp$ limit.

The polarisation equation \cite{wwlee} is given by Eqs.\ (7,48),
rewritten as Eq.\ (93), all from Ref.\ \cite{beer},
\begin{equation}
\nefl = {\nifl\over 1+b/2}-{(b/2)\tippfl\over(1+b/2)^2}
	+{\Gamma_0-1\over\tau_i}\phifl
\end{equation}
preserving the normalisation of $\phifl$
in terms of $T_e$, where $\tippfl$ is the
perpendicular temperature disturbance defined as the normalised
$(\vpp^2-1)$ moment of the perturbed distribution function.
Hence three gyroaveraging operators appear in the moment equations, but
only two appear in the polarisation.  Moreover, the Pad\'e approximate
forms are used in the latter but not in the former.  Even if this were
repaired, using
\begin{equation}
\Gamma_0^{1/2} \to {1\over 1+b/2}
\end{equation}
in the gyroaveraging of the potential (cf.\ Sec.\ III.C.4 of Ref.\
\cite{dorland}), 
it would be impossible to reconcile the fact that the $\ddppgg$ operator
does not appear in the polarisation equation, which in terms of the
gyroaveraging operators reads
\begin{equation}
\nefl=\Gamma_1\nifl+\Gamma_2\tippfl +{\Gamma_0-1\over\tau_i}\phifl
\label{eqbpol}
\end{equation}
For the purposes of energetic consistency it does not matter how or
whether the $\Gamma_0-1$ screening term is approximated, only which
operators appear in the $\nifl$ and $\tifl$ terms.

The ExB energy is given by 
\begin{equation}
\int d\Lambda\,{1-\Gamma_0\over\tau_i}{\phifl^2\over 2}
\end{equation}
Using Eq.\ (\ref{eqbpol}) we may replace this according to
\begin{equation}
\int d\Lambda\LP
	{1-\Gamma_0\over\tau_i}{\phifl^2\over 2}+{\phifl\nefl\over 2}\RP
= \int d\Lambda\,\phifl\LP
	{\Gamma_1\nifl+\Gamma_2\tippfl}\over 2\RP
\end{equation}
Using the Hermitian property of the gyroaveraging operators and the
definitions of the potentials, and placing the electron contribution on
the right side, we find
\begin{equation}
\int d\Lambda\,
	{1-\Gamma_0\over\tau_i}{\phifl^2\over 2}
= \int d\Lambda\,\LP
	{\phigfl\nifl+\vorgfl\tippfl\over 2}-{\phifl\nefl\over 2}\RP
\end{equation}
The drift energy is therefore expressed as a combination of potentials
multiplied by thermal state variables.

If the electrons are adiabatic due to fast parallel dynamics on closed
flux surfaces, the electron density is itself replaced by the potential
according to
\begin{equation}
\nefl=\phifl-\avg{\phifl}
\end{equation}
where the angle brackets denote the flux surface (``zonal'') average.
In this case the electron contribution is a sort of field energy which
is combined with the proper drift energy to form the potential energy
given by
\begin{equation}
\int d\Lambda\,\phifl\LP{\phifl-\avg{\phifl}\over 2}+
	{1-\Gamma_0\over\tau_i}{\phifl\over 2}\RP
= \int d\Lambda\,\LP
	{\phigfl\nifl+\vorgfl\tippfl\over 2}\RP
\end{equation}
The flux surface average is subtracted because
the adiabatic state arises through the large value of the parallel
wavenumber $\kpl$ combined with the electron thermal velocity, in
comparison with the dynamical frequencies, while for the zonal component
there is no action by $\dpl$.
The adiabatic electron approximation does not hold in general,
but it is assumed in Refs.\ \cite{dorland,beer}, and keeping it within
this Section makes this discussion more transparent.

The thermal free energy is given by the fluid moment variables in
quadratic combination.  The state variable free energy is given by
\begin{equation}
\int d\Lambda\,\LP\tau_i
	{\nifl^2+\tippfl^2+\tiplfl^2\over 2}\RP
\end{equation}
while the flux variable free energy (the ``generalised parallel kinetic
energy'') is given by
\begin{equation}
\int d\Lambda\,\LP\epss
	{\ufl^2+\qippfl^2+\twothirds\qiplfl^2\over 2}\RP
\end{equation}
Comparison of all these pieces leads to the conclusion that the
combinations $\tau_i\nifl+\phigfl$ and $\tau_i\tippfl+\vorgfl$ should
appear together under first derivatives in linear terms in the 
moment equations in order that in combination among all the energy
pieces the various terms reduce to terms involving 
single first derivative operators
acting upon combinations of the variables, which have the form of total
divergences, either $B\dpl$ or $\kappacv$.  It is clear that since no
operation by the third gyroaverage $\Gamma_3$ appears in the
polarisation equation with only densities and temperatures kept in the
closure for the total space density, there is no place for the third
gyroreduced potential $\vorggfl$ in the moment equations.  This is
the first and most obvious inconsistency of the standard gyrofluid
model, and it impacts the parallel dynamics, the curvature terms
(quasistatic compressible part of the drift dynamics), and the magnetic
pumping process, each of which we presently examine in turn.  Finally,
the additional inconsistency in the treatment of higher moments in the
curvature terms in the equations for $\qiplfl$ and $\qippfl$ is addressed.

We first look at the parallel dynamics.  This involves conservative
energy transfer in the sound waves and conductive heat fluxes.  The
simplest case of a sound wave in the absence of any effects due to the
potential is of a parallel gradient in the pressure causing a parallel
flow, and the corresponding parallel compression of that flow acting to
restore the pressure disturbance.  The total divergence term expressing
energy conservation in such a local model as this one is
$B\dpl(\pfl\ufl/B)$, up to numerical constant factors dependent on how
the temperatures are described.  This divergence represents a transport
process, in this case advection of thermal energy by $\ufl$.
When the potential is also involved, charge currents become part of
this, but the structure is the same.  

In the standard gyrofluid model,
the part of the dynamics involving sound waves is
\begin{eqnarray}
\ptt{\nifl}=-B\dpl{\ufl\over B}\\
\epss\ptt{\ufl}=-\dpl\LB\tau_i\LP\nifl+\tiplfl\RP+\phigfl\RB\\
\half\ptt{\tiplfl}=-B\dpl{\ufl\over B}
\end{eqnarray}
in addition to the polarisation equation.
Evolution of the potential energy, the
thermal state variable energy, and the thermal flux variable energy
pieces involved in this is given by
\begin{eqnarray}
\phigfl\ptt{\nifl}=-B\dpl{\phigfl\ufl\over B}+\ufl\dpl\phigfl\\
\epss\ufl\ptt{\ufl}=-\ufl\dpl\LB\tau_i\LP\nifl+\tiplfl\RP+\phigfl\RB\\
\tau_i\nifl\ptt{\nifl}=-B\dpl{\tau_i\nifl\ufl\over B}
	+\ufl\dpl\tau_i\nifl\\
\half\tau_i\tiplfl\ptt{\tiplfl}=-B\dpl{\tau_i\tiplfl\ufl\over B}
	+\ufl\dpl\tau_i\tiplfl
\end{eqnarray}
When the pieces are summed, the transfer terms denoted by $\ufl\dpl$
cancel, leaving the total transport divergence term denoted by
\begin{equation}
B\dpl{\LB\tau_i\LP\nifl+\tiplfl\RP+\phigfl\RB\ufl\over B}
\end{equation}
that is, just the divergence of a transport flux given by
the force potential in the equation for $\ufl$
multiplied by $\ufl\bb/B$.
Here we note that $\tippfl$ is not involved, so there is no action by
$\vorgfl$.  We find by this analysis that the sound wave dynamics is in
order and does not require modification.  The same conclusion results
from examination of the non-closure part of the heat conduction
dynamics, for which the equation parts are
\begin{eqnarray}
\half\ptt{\tiplfl}=-B\dpl{\qiplfl\over B}\\
\epss\ptt{\qiplfl}=-\threehalves\dpl\LB\tau_i\tiplfl\RB
\end{eqnarray}
and
\begin{eqnarray}
\ptt{\tippfl}=-B\dpl{\qippfl\over B}\\
\epss\ptt{\qippfl}=-\dpl\LB\tau_i\tippfl+\vorgfl\RB
\end{eqnarray}
and the energetics parts are
\begin{eqnarray}
\half\tau_i\tiplfl\ptt{\tiplfl}=-B\dpl{\tau_i\tiplfl\qiplfl\over B}
	+ \qiplfl\dpl\LB\tau_i\tiplfl\RB\\
\twothirds\epss\qiplfl\ptt{\qiplfl}=
	-\qiplfl\dpl\LB\tau_i\tiplfl\RB
\end{eqnarray}
and
\begin{eqnarray}
\tau_i\tippfl\ptt{\tippfl}=
	-B\dpl{\LB\tau_i\tippfl+\vorgfl\RB\qippfl\over B}
	+ \qippfl\dpl\LB\tau_i\tippfl+\vorgfl\RB\\
\epss\qippfl\ptt{\qippfl}=
	-\qippfl\dpl\LB\tau_i\tippfl+\vorgfl\RB
\end{eqnarray}
Here we note the factor of two difference in the definition of $\qiplfl$
here (conformal with the Braginskii definition \cite{brag}) and in the
standard model. 

There are also magnetic pumping terms in the gyrofluid parallel
dynamics, due to the combination of parallel flow and conduction,
magnetic moment conservation at the gyrokinetic level, and the parallel
gradient in the strength of the magnetic field.  Here, the standard
model has the terms in the right places except for a single occurrence
of the ``forbidden'' potential $\vorggfl$:
\begin{eqnarray}
\epss\ptt{\ufl}=-\LB\tau_i\LP\tippfl-\tiplfl\RP+\vorgfl\RB\dpl\log B\\
\half\ptt{\tiplfl}=-\LP\qippfl+\ufl\RP\dpl\log B\\
\ptt{\tippfl}=\LP\qippfl+\ufl\RP\dpl\log B\\
\epss\ptt{\qippfl}=-\LB\tau_i\LP\tippfl-\tiplfl\RP
	+2\vorggfl-\vorgfl\RB\dpl\log B
\end{eqnarray}
If we merely replace $\vorggfl$ by $\vorgfl$ we restore consistency,
\begin{equation}
\epss\ptt{\qippfl}=-\LB\tau_i\LP\tippfl-\tiplfl\RP+\vorgfl\RB\dpl\log B
\end{equation}
When the free energy pieces are constructed, these effects form 
transfer channels which then properly conserve energy because
$\tau_i\tippfl$ and $\vorgfl$ occur in combination.  The reason
$\vorggfl$ is ``forbidden'' is that it doesn't appear in the
polarisation equation.  If it is present, then the conservation in the
exchange between potential and thermal energy is broken.

A similar problem appears in the curvature terms.  For the thermal state
variables in the standard model, these are
\begin{eqnarray}
\ptt{\nifl}=\kappacv\LP\phigfl+{\vorgfl\over 2}
	+\tau_i{\piplfl+\pippfl\over 2}\RP\\
\half\ptt{\tiplfl}=\kappacv\LP{\phigfl+\tau_i\piplfl\over 2}
	+\tau_i\tiplfl\RP\\
\ptt{\tippfl}=\kappacv\LP{\phigfl+\vorgfl+\tau_i\pippfl\over 2}
	+{3\tau_i\tippfl+\vorgfl+2\vorggfl\over 2}\RP
\end{eqnarray}
Again, we need merely replace $\vorggfl$ by $\vorgfl$ to restore
consistency,
\begin{equation}
\ptt{\tippfl}=\kappacv\LP{\phigfl+\vorgfl+\tau_i\pippfl\over 2}
	+3{\tau_i\tippfl+\vorgfl\over 2}\RP
\end{equation}
so that $\tau_i\tippfl$ and $\vorgfl$ again occur in combination.  Once
more, thermal free energy was already conserved in the standard model,
but due to $\vorggfl$ a mismatch in the FLR part of the transfer between
potential energy and thermal free energy remained.

The curvature terms in the fluid moment flux variables present a
different problem, the only inconsistency in the standard model which is
not a FLR effect.  There are no curvature terms involving the potential
in the equations for $\ufl$, $\qiplfl$, and $\qippfl$, but there is a
closure treatment at the level of the fifth moments which appears in the
equations for the third moments ($\qiplfl$ and $\qippfl$), arising from
the factors of $\vpp^2$ and $\vpl^2$ in the grad-B and curvature drift
terms in the gyrokinetic equation.  In the standard model the closure
for the 4th and 5th moments is taken from a perturbed Maxwellian (cf.\
its Eqs.\ 81 and 82).  However, the model retains $\qiplfl$ and
$\qippfl$ as dynamical variables, and so the 5th moment should include
contributions from pressures times conductive heat fluxes.  For example,
parts of the $\vpp^2\vpl^3$ moment is provided under drift ordering by
$\pipp\qiplfl$ and $\pipl\qippfl$,
and part of the $\vpp^4\vpl$ moment is provided by
$\pipp\qippfl$.  In the normalisation, the factors of $\pipl$ and
$\pipp$ are replaced by unity.  A way to do this systematically is to
express the perturbed
distribution function as a general six-term polynomial in
which each of the coefficients is represented by one of the fluid moment
variables retained in the six-moment model.  Then, the fifth moments are
computed by evaluating the integrals over $\vpp^2\vpl$ or $\vpl^3$ times
the perturbed
distribution function.  The result of this calculation is a combination
which automatically conserves free energy within the fluid moment
system:
\begin{eqnarray}
\epss\ptt{\ufl}={\epss\tau_i\over 2}\kappacv\LP
	4\ufl+2\qiplfl+\qippfl\RP\\
\epss\ptt{\qiplfl}={\epss\tau_i\over 2}\kappacv\LP
	3\ufl+8\qiplfl\RP\\
\epss\ptt{\qippfl}={\epss\tau_i\over 2}\kappacv\LP
	\ufl+6\qippfl\RP
\end{eqnarray}
The ``diagonal'' terms in the implied curvature matrix conserve
automatically, but what this procedure has done is to ensure that the
``cross'' terms also act conservatively.

The final consideration is the ExB advection terms, especially their FLR
generalisations.  All three potentials ($\phigfl$, $\vorgfl$, and
$\vorggfl$) are involved in the standard model, acting through their
respective drift velocities,
\begin{equation}
\uexb=-\fdel\phigfl \qquad
\wexb=-\fdel\vorgfl \qquad
\Wexb=-\fdel\vorggfl
\end{equation}
where the tensor operator $-\fdel$ represents the more familiar
$(c/B^2)\bb\cross\grad$.  The standard model has ``diagonal'' terms, in
which each variable is acted upon by $\uedl$ in its own equation,
``cross'' terms in which the pairs $(\nifl,\tippfl)$ and
$(\ufl,\qippfl)$ are coupled by $\wedl$, and then extra ``diagonal'' FLR
effects in which $\tippfl$ and $\qippfl$ are acted upon by $2\Wedl$ in
their own equations.  Because $\vorggfl$ is not $\vorgfl$, the FLR
potential energy $\vorgfl\tippfl/2$ is not conserved by $\Wedl\tippfl$.
Again, this is repaired by simply replacing $\vorggfl$ by $\vorgfl$.

Two minor considerations remain: first, the part of the collision
process which enforces isotropisation should have the collision
frequency multiplying $\tippfl+\vorgfl-\tiplfl$ rather than simply
$\tippfl-\tiplfl$, so that the total dissipation is positive definite.
This merely captures the correspondence between a gyrofluid
$\tippfl+\vorgfl$ and a fluid $\tipp$ as discussed by Belova
\cite{belova}.
Second, we have discussed energetics independently of nondissipative
closure.  The standard model constructs a curvature dissipation matrix
in order to capture toroidal drift phase mixing by the
velocity-dependent grad-B and curvature drifts.  Part of this is
nondissipative and can in principle capture the 5th moment effects
discussed above.  However, we choose here to separate these effects
because in some applications involving nonperiodic drifts there should
be no phase mixing effect at all.  Moreover, the dissipation matrix
treatment itself is not a real success.  It is explained (pp.\ 4057-8 of
Ref.\ \cite{beer}) that a different set of coefficients was required to
accurately represent the kinetic result close to marginal stability for
the adiabatic-electron, toroidal ITG mode.  This is in itself an
admission of failure for the project of using the gyrofluid system as a
quantitatively exact representation of the gyrokinetic one.  We do not
attempt such an ambitious goal here; rather, the GEM model is intended
for qualitative study of basic physics mechanisms, especially energy
transfer between small scale turbulence and large scale flows and MHD
processes.  Hence the neglect of dissipation free closure effects, and
in general the neglect of dissipative effects other than collisions and
Landau damping.

The main point of this Section has been to highlight the way in which
potential energy and thermal free energy are conservatively exchanged by
the various mechanisms involved in low frequency fluid drift dynamics
and their capture by a gyrofluid model which is at least well behaved to
arbitrary order in the finite Larmor radius parameter $\kpp\rho_i$.  If
a turbulence 
model is to act at arbitrary FLR order it should conserve energy
properly, even if for no other reason than that a numerical computation
should not experience 
trouble in the spectral region around $\kpp\rho_i\sim 1$.  
The procedure to repair the standard model accordingly is to restore the
use of the same
closure treatment (and to the same depth in the number of moments kept)
in the polarisation equation(s) and in the fluid moment equations.  In
the end, the same gyroaveraging operators appear in the polarisation and
fluid moment equations --- since the polarisation equation in the
six-moment model only involves $\nifl$ and $\tippfl$ it only involves
the first two operators $\Gamma_1$ and $\Gamma_2$, and hence only two
potentials (one nominal, $\phigfl$, and one FLR, $\vorgfl$) can appear
in the fluid moment equations.
The third gyroaverage operator $\Gamma_3$ may have a reasonable role in
a gyrofluid model which retains 4th and 5th moments (one level in the
hierarchy past temperatures and conductive heat fluxes), but we do not
pursue this extension herein.

}

\section{Construction of the GEM Model
\label{sec:gem}}

Subject to the conventions in Section \ref{sec:norm},
our starting point is the polarisation equation, which links the
variables $\nzfl$ and $\tzppfl$ for each species to the electrostatic
potential, $\phifl$.  We neglect true space charge effects, setting the
Debye length to zero and assuming the space charge densities all add up
to zero.  This is the statement of quasineutrality:
\begin{equation}
\sum_z a_z \LB \Gamma_1\nzfl+\Gamma_2\tzppfl
+{\Gamma_0-1\over\tau_z}\phifl\RB = 0
\label{eqpolarisationz}
\end{equation}
where the gyroaveraging and screening operators are defined separately
for each species,
\begin{equation}
\Gamma_0 = \Gamma_0(b_z)\qquad
\Gamma_1 = \Gamma_0^{1/2}(b_z)\qquad
\Gamma_2 = b_z\pbb{\Gamma_1}(b_z)
\end{equation}
with argument $b_z=\kkpp\rho_z^2$ and squared gyroradius
$\rho_z^2=\mu_z\tau_z/B^2$, where $B$ is the normalised strength of the
equilibrium magnetic field.
The considerations which lead to these forms are the ones given in the
standard gyrofluid model \cite{dorland}, as outlined in Section
\ref{sec:standard}.

We identify the generalised ExB energy using the polarisation densities
in Eq.\ \ref{eqpolarisationz} as
\begin{equation}
U_E = \sum_z a_z {\Gamma_0-1\over\tau_z}{\phifl^2\over 2}
\end{equation}
Using Eq.\ \ref{eqpolarisationz} and the Hermitian property of the
$\Gamma$ operators, we recast this as
\begin{equation}
U_E = \sum_z a_z {\phigfl\nzfl+\vorgfl\tzppfl\over 2}
\end{equation}
where the gyroreduced potentials are given by
\begin{equation}
\phigfl = \Gamma_1\phifl \qquad\qquad \vorgfl = \Gamma_2\phifl
\label{eqpotentials}
\end{equation}
Note that these are defined separately for each species and that there
is no $\Gamma_3$.

We identify the thermal state variable part of the energy the same way
as in the fluid models,
\begin{equation}
U_t = \sum_z a_z\tau_z{\nzfl^2+(1/2)\tzplfl^2+\tzppfl^2\over 2}
\end{equation}
The flux variable part of the energy is
\begin{equation}
U_v = \sum_z a_z\mu_z{\uzfl^2+(2/3)\qzplfl^2+\qzppfl^2\over 2}
\end{equation}
We note here that since the evolution of $U_t$ and $U_E$ ultimately
follows from the same moment equations, the combinations which must
appear together under the $\dpl$ and $\kappacv$ operators in those
equations are $\phigfl+\tau_z\nzfl$ and $\vorgfl+\tau_z\tzppfl$.
Observing this will guarantee energetic consistency.

The magnetic field disturbances arise from the parallel magnetic
potential, which is given by Ampere's law in terms of the total electric
current,
\begin{equation}
-\ddpp\Afl = \Jfl = \sum_z a_z \uzfl
\label{eqampere}
\end{equation}
The magnetic energy is given by 
\begin{equation}
U_m = {\beta_e\over 2}\abs{\dpp\Afl}^2
\end{equation}
which using Eq.\ (\ref{eqampere}) may be rewritten as
\begin{equation}
U_m = \beta_e{\Afl\Jfl\over 2}
\label{eqmagenergy}
\end{equation}
This model neglects gyroaveraging and gyroscreening of the
magnetic potential.
On the same footing as the potential equation we would have a more
general Ampere's law in which $\uzfl$ might be replaced by 
$\Gamma_1\uzfl+\Gamma_2\qzppfl$, and due to the same consistency
considerations as with $\phigfl$ and $\vorgfl$ several new finite
gyroradius terms would appear in the magnetic flutter dynamics.
This is being left for future work, however, because the consequences
for energy conservation have not yet been worked out.

In situations wherein the finite electron gyroradius is neglected, we
simply have $\phigfl=\phifl$ and $\vorgfl=0$ for the electrons, so that
the polarisation equation becomes
\begin{equation}
\sum_i a_i \LB \Gamma_1\nifl+\Gamma_2\tippfl
+{\Gamma_0-1\over\tau_i}\phifl\RB = \nefl
\label{eqpolarisationi}
\end{equation}
where the species label is changed to $i$ as it refers to the ions
only.  If as in most practical applications one takes a single component
plasma with singly charged ions, we merely have $a_i=1$ and
$\tau_i=T_i/T_e$, and with normalisation to that particular ion's mass,
$\mu_i=1$ and $\mu_e=-m_e/M_i$ along with $a_e=\tau_e=-1$.  Since these
are trivial restrictions, we present the GEM model in terms of the 
general forms following from Eq.\ (\ref{eqpolarisationz}).

The ExB advection operators and the nonlinear part of the parallel
gradient are given in terms of Poisson bracket structures,
\begin{equation}
\uedl=[\phigfl,]\qquad\qquad
\wedl=[\vorgfl,]\qquad\qquad
\bbdl=-\beta_e[\Afl,]
\end{equation}
in the two perpendicular coordinates, that is,
\begin{equation}
[f,g]=\LP\pxx{f}\pyy{g}-\pxx{g}\pyy{f}\RP
\end{equation}
Defined in this fashion, the
quantities $\uexb$, $\wexb$, and $\bperp$ are all divergence free as
written; the generally finite divergences of $\uexb$ and $\wexb$ are
treated separately, through $\kappacv(\phigfl)$ and $\kappacv(\vorgfl)$,
respectively.  The advective time derivative and the parallel derivative
are given by
\begin{equation}
\dtt{}=\ptt{}+\uedl\qquad\qquad
\dpl = {1\over B}\Bdel + \bbdl
\end{equation}
where $\bb$ and $B$ are defined in terms of the equilibrium magnetic
field.  The variation of $B$ along a field line (poloidally around the
flux surface in a tokamak) is incorporated into the gyroradii
\begin{equation}
\rho_z^2 = {\mu_z\tau_z\over B^2}
\end{equation}
The perpendicular parts of the Laplacian are given by
\begin{equation}
\ddpp = {1\over g^{1/2}}\pxxmu{}g^{1/2}\,g^{\mu\nu}_\perp\pxxnu{}
\end{equation}
where $g$ is the determinant of the $\{g_{\mu\nu}\}$ elements of the
entire metric, and 
$g^{\mu\nu}_\perp$
is the perpendicular metric which involves only the two perpendicular
coordinates ($x$ and $y$).
In these terms, the argument $b_z$
appearing in the $\Gamma$ operators is given by
\begin{equation}
b_z = \rho_z^2 \LP k_\mu g^{\mu\nu}_\perp k_\nu \RP
\end{equation}
where $\mu$ and $\nu$ are summed over the two perpendicular coordinates
only. 
Note that $\rho_z^2$ commutes with $\ddpp$ and that
no gyroradius
appears with the $\ddpp$ operator in the Ampere's law
(Eq.\ \ref{eqampere}).  

\newpage

The six moment equations are the same ones appearing in the toroidal
version of the standard gyrofluid model \cite{beer}, corrected according
to the findings in Section \ref{sec:standard}.  With $\nzfl$ and
$\phigfl$, and $\tzppfl$ and $\vorgfl$, appearing together, and the
background gradient forcing terms $n_z$ and $T_z$ (functions of $x$
only) displayed explicitly, the dissipation free part of the equations
for each species are
\begin{equation}
\dtt{\Nzfl}+\wedl\Tzppfl+B\dpl{\uzfl\over B}
	= \kappacv\LP\phigfl+{\tau_z\pzplfl+\tau_z\pzppfl+\vorgfl\over 2}\RP
\label{eqnzfl}
\end{equation}
\begin{eqnarray}
&& \beta_e\ptt{\Afl}+\mu_z\dtt{\uzfl} + \mu_z\wedl\qzppfl 
	= - \dpl\LP\phigfl+\tau_z\Pzplfl\RP 
	\nonumber\\ && \qquad\qquad{}
	+ \kappacv\LP\mu_z\tau_z{4\uzfl+2\qzplfl+\qzppfl\over 2}\RP
	- \tau_z\LP\vorgfl+\tau_z\tzppfl-\tau_z\tzplfl\RP\dpl\log B
\label{equzfl}
\end{eqnarray}
\begin{eqnarray}
&&
\half\dtt{\Tzplfl}+B\dpl{\uzfl+\qzplfl\over B}
	\nonumber\\ && \qquad\qquad{}
	= \kappacv\LP{\phigfl+\tau_z\pzplfl\over 2} + \tau_z\tzplfl\RP
	- \LP\uzfl+\qzppfl\RP\dpl\log B
\label{eqtzplfl}
\end{eqnarray}
\begin{eqnarray}
&& \dtt{\Tzppfl}+\wedl\LP\Nzfl+2\Tzppfl\RP+B\dpl{\qzppfl\over B}
	\nonumber\\ && \qquad\qquad{}
	= \kappacv\LP{\phigfl+\vorgfl+\tau_z\pzppfl\over 2}
		+3{\vorgfl+\tau_z\tzppfl\over 2}\RP
	+ \LP\uzfl+\qzppfl\RP\dpl\log B
\label{eqtzppfl}
\end{eqnarray}
\begin{eqnarray}
\mu_z\dtt{\qzplfl}
	= - \threehalves\dpl\LP\tau_z\Tzplfl\RP
	+ \kappacv\LP\mu_z\tau_z{3\uzfl+8\qzplfl\over 2}\RP
\label{eqqzplfl}
\end{eqnarray}
\begin{eqnarray}
&& \mu_z\dtt{\qzppfl} + \mu_z\wedl\LP\uzfl+2\qzppfl\RP
	= - \dpl\LP\vorgfl+\tau_z\Tzppfl\RP
	\nonumber\\ && \qquad\qquad{}
	+ \kappacv\LP\mu_z\tau_z{\uzfl+6\qzppfl\over 2}\RP
	- \tau_z\LP\vorgfl+\tau_z\tzppfl-\tau_z\tzplfl\RP\dpl\log B
\label{eqqzppfl}
\end{eqnarray}
The pressures are defined as
\begin{equation}
\pzplfl=\nzfl+\tzplfl\qquad\qquad
\pzppfl=\nzfl+\tzppfl\qquad\qquad
p_z = n_z + T_z
\end{equation}
and all the thermal state variables are operated upon by $\uedl$,
$\wedl$, and $\dpl$ together with their gradients.

\newpage

\subsection{Local and Global Models}

The GEM model is variously cast in both local and global versions.  In
the local version the gradient drive terms appear explicitly in the
equations as displayed above.  Following the ordering, the pressures add
linearly also in these quantities.
The densities and temperatures are given prescribed forms, usually simple
linear gradients, e.g., $n_z=-\wn\,x$ such
that $\wn=\abs{\Lpp\grad\log n_z}$ gives the inverse of the
normalised scale
length, but they can be given arbitrary form, allowing computations
within any prescribed gradients.  
For this version all dependent
variables are given Dirichlet boundary conditions in the $x$-direction,
\begin{equation}
  f = 0 \quad\hbox{at}\quad x=\pm{L_x\over 2}
\end{equation}
where $L_x$ is the domain length.

In the global version the separate gradient terms do not appear in the
moment equations.  Instead, the fluctuating gradient is part of the
dependent variable.  Accordingly, the dependent variables are given
Neumann/Dirichlet boundary conditions in the $x$-direction,
\begin{equation}
\pxx{f} = 0 \quad\hbox{at}\quad x=-{L_x\over 2} \qquad\qquad
  f = 0 \quad\hbox{at}\quad x={L_x\over 2}
\end{equation}
allowing arbitrary profile evolution.  In this version the curvature
operator and also all the dissipation operators act upon the entire
variable, including the profile.  Consequently, the two dimensional
equilibrium including parallel flows and currents, and heat fluxes,
is also solved for and evolved self consistently with the
turbulence.  The transport problem is also solved.  One usually operates
in one of two limits: the run time is either longer or much shorter than
the confinement time.  In the latter case sources are not necessary; the
profile is allowed to relax but is expected not to do so very much
(perhaps $30\%$ relaxation is acceptable).  For edge turbulence in a
thin radial layer such that $L_x<\Lpp$, the confinement time is usually
shorter than the time for the zonal flows, resulting from the evolving
flux surface (``zonal'') averaged potential, to reach statistical
equilibrium.  In this case sources are necessary for all the state
variables ($\nzfl$ and $\tzplfl$ and $\tzppfl$ for each species).  In
either global
case, the profiles cannot be prescribed except for the fact that
the system is initialised with the profiles set into the state variables.

A final consideration in the global model is polarisation: vorticity is
given in general by the ``gyrocenter charge density'' made up by $\sum_z
a_z \nzfl$, noting that it is the total charge density that is set to
zero.  Here, the profiles are included in the dependent variables, so
especially in the adiabatic electron version one must keep the
unchanging $\avg{\nefl}$ in polarisation,
\begin{equation}
\nefl = \avg{\nefl} + \phifl - \avg{\phifl}
\end{equation}
where the angle brackets denote the zonal average.  In the local
version, the profile function $n_i(x)$ does not appear in polarisation,
since it is expected to be equal to $n_e(x)$.

Both global and local versions are set up with globally consistent
boundary conditions parallel to the background magnetic field.
This ensures individual fulfillment of the periodicity constraint
for each Fourier component in $y$ as if on the entire flux surface,
even if the toroidal mode spectrum is truncated \cite{fluxtube}.
Here we note
that $k_y$ follows the toroidal mode number generally and hence the
$y$-domain is periodic.  The domain length for $y$
is $L_y$.  The periodicity constraint applies as a boundary condition in
$s$ after one single poloidal cycle, so the $s$-domain is always one
connection length $-\pi<s<\pi$.
Additionally, the $y$-coordinate is shifted on each drift plane
(constant-$s$ surface), so that perpendicular dynamics is always
computed with an orthogonal metric, and magnetic shear enters as a set
of relative shifts in the $y$-coordinate in the expressions for
$\pps{}$, as explained in Ref.\ \cite{shifted}.  This combination
is required for capture of slab-character modes, of which the most
important in turbulence is the nonlinear drift wave instability
\cite{ssdw,focus,eps03}.  Global consistency is also required to obtain
the correct spectrum of sideband modes in the equilibrium, to which the
turbulence and zonal flows are coupled by toroidal compression of the
ExB velocity \cite{gdcurv}.

\subsection{Fourier and Pad\'e Versions}

The operators involving $\ddpp$, including the $\Gamma$'s, are solved
variously in $xy$-space or in $\kperp$-space.  The Fourier versions are
set up to be compatible with either the local or global boundary
conditions.  The local model with Dirichlet boundaries uses half-wave
Fourier transforms, with the basic $x$-domain odd-reflected about
$x=L_x/2$, that is, $f(L_x/2+x')=-f(L_x/2-x')$ for $0<x'<L_x$.  The
Fourier transform and its inverse is then applied to the doubled domain.
In a similar manner, the global model uses quarter-wave Fourier
transforms, with four copies of the $x$-domain arrayed
odd-even-even-odd, and with the transforms applied to the quadrupled
domain.  This Fourier version of either the local or global model is
used whenever the argument $b_z=\kkpp\rho_z^2$ of any species is
expected to take large values.  The hallmark example of this is ETG
turbulence (Sec.\ \ref{sec:results}), which involves the entire scale
range between $\rho_i$ and $\rho_e$ in a single component plasma.  If
the ion moment variables are kept, the Fourier version should be used to
obtain an accurate ion response, which becomes more and more
``adiabatic'' (in the sense that $\phifl\to-\tau_i\nefl$) with
increasing $b_i$.  In current implementations, the Fourier version
always uses the full FLR form for every species, including electrons,
regardless of the expected values of $b_z$.

For standard ITG or edge turbulence cases, where the scale range reaches
down to but not below $\rho_i$, the Pad\'e version may be used.  This
approximates the $\Gamma$'s by \cite{dorland}
\begin{equation}
\Gamma_0(b_z) \to \LP 1-\rho_z^2\ddpp\RP^{-1}
\end{equation}
\begin{equation}
\Gamma_1(b_z) \to \LP 1-\half\rho_z^2\ddpp\RP^{-1}
\end{equation}
\begin{equation}
\Gamma_2(b_z)=\half\rho_z^2\ddpp\LP 1-\half\rho_z^2\ddpp\RP^{-1}\Gamma_1(b_z)
\end{equation}
However, if the finite gyroradius
effects of more than one species are taken into account, one must solve
the combined screening operator given by 
$\sum_z (a_z/\tau_z)(\Gamma_0-1)$ to find $\phifl$.  This is simple in
$\kperp$-space but complicated if using the Pad\'e forms in
configuration space, even if one of the coordinates is Fourier
decomposed.  With only two gyroradii to follow, however, one has the
acceptable operation involving two successive Helmholtz solves.
In current implementations, the Pad\'e version is only used for single
component plasma cases in which the electron FLR effects are neglected.
The adiabatic ion model is only used with the Pad\'e version.  
The importance of the Pad\'e version is that it is the only one easily
generalised to fully global geometry, wherein the metric coefficients
depend on $x$.

\subsection{Dissipation in GEM -- Collisionless
\label{sec:dissip}}

The only true dissipation in the collisionless GEM model is phase mixing
due to the kinetic resonances caused by the parallel transit dynamics,
i.e., Landau damping.  This
is represented by direct dissipation upon the parallel heat flux
variables, using a Landau damping operator defined by
\begin{equation}
\aldz\equiv\ald\LP 1 - 0.125 V qR\ddpl \RP
\end{equation}
with constant $\ald$ nominally set to unity,
where $qR$ is the field line connection length divided by $2\pi$
\cite{fluxtube}, $V={\tau_z/\mu_z}$ is the normalised thermal speed
of species $z$, and 
$\ddpl$ is generally the full nonlinear parallel Laplacian
divergence operator $B\dpl(1/B)\dpl$.
Both $qR$ and $\ddpl$ are normalised in terms of $\Lpp$.

The GEM model does not employ a curvature drift dissipation model.  The
standard one did so \cite{beer}, but also noted in detail the problems
it raised.  For the reasons discussed in Section \ref{sec:standard},
therefore, it is chosen to omit this feature.  Nonlinear FLR phase
mixing \cite{dorland} is also left out of the standard model
\cite{beer}, for similar reasons of tractability.

\subsection{Dissipation in GEM -- Collisional
\label{sec:coll}}

The electromagnetic gyrofluid model finds a very useful application in
tokamak edge turbulence \cite{gyroem}, and hence requires a treatment of
collisions which will capture the collisional Braginskii fluid model
\cite{brag} in the appropriate limit of large collision frequency and
short mean free path.  The usual types of dissipation are resistivity
and thermal conduction, which in a model treating both velocities and
heat fluxes as dynamical variables amounts to applying a dissipation
matrix to their combination.  For an isotropic temperature the method
used in the DALF Landau fluid model \cite{dalfloc} is sufficient.  The
gyrofluid model, however, additionally includes temperature anisotropy.
To combine these effects, a simple drift kinetic Chapman-Enskog
procedure is used to find the dissipative corrections to a state with
stationary state variables
\cite{hassam}, generalised to a bi-Maxwellian distribution (defined by
$\nzfl$, $\tzplfl$, and $\tzppfl$ for each species) for the dissipation
free part.  A Lorentz collision operator is used, and then the
Braginskii coefficients are substituted to capture the collisional limit.
The resulting model is given by
\begin{equation}
\beta_e\ptt{\Afl}+\mu_z\dtt{\uzfl} 
	= \cdots + \mu_e\nu_e \LB \eta\Jfl
		+ {\alpha_e\over\kappa_e}\LP \qeplfl+\qeppfl+\alpha_e\Jfl\RP\RB
\end{equation}
\begin{equation}
\half\dtt{\tzplfl}
	= \cdots - \nu_z\LB\tau_z\LP\tzplfl-\tzppfl\RP-\vorgfl\RB
\end{equation}
\begin{equation}
\dtt{\tzppfl}
	= \cdots + \nu_z\LB\tau_z\LP\tzplfl-\tzppfl\RP-\vorgfl\RB
\end{equation}
\begin{equation}
\mu_z\dtt{\qzplfl}
	= \cdots - {(5/2)\over\kappa_z}\mu_z\nu_z
		\LP\qzplfl-0.6\alpha_z\Jfl\RP 
		+ 1.28\nu_z\LP\qzplfl-1.5\qzppfl\RP
\end{equation}
\begin{equation}
\mu_z\dtt{\qzppfl}
	= \cdots - {(5/2)\over\kappa_z}\mu_z\nu_z
		\LP\qzppfl-0.4\alpha_z\Jfl\RP 
		- 1.28\nu_z\LP\qzplfl-1.5\qzppfl\RP
\end{equation}
where $\kappa_z$ and $\alpha_z$ are the thermal conduction and thermal
force coefficients for each species, and $\npl$ is the resistivity
coefficient.  Herein, the thermal force is kept only for electrons
($\alpha_z=\alpha_e$), while for ions it is zero.  For a single
component plasma with singly charged ions, the values of the
coefficients are 
\begin{equation}
\eta=0.51 \qquad \alpha_e=0.71 \qquad \kappa_e=3.2
	\qquad \kappa_i=3.9
\end{equation}
To recover an
isotropic model we would add $(1/2)\tzplfl+\tzppfl$ to form $(3/2)\tzfl$
and $\qzplfl+\qzppfl$ to form $\qzfl$.  This yields the forms in Ref.\
\cite{dalfloc}.  Then, we could recover the Braginskii formula for
$\qzfl$ by neglecting the inertial and Landau damping terms in its
equation (i.e., neglect $\qzfl$ except in the collisional damping term),
as explained in Ref.\ \cite{dalfloc}.

The terms in the temperature equations and the ones with the factors of
$1.28$ represent relaxation of anisotropy, and the others represent
resistive $(\eta)$ and thermal conductive $(\kappa_z)$ dissipation.
Note the combination $\tau_z\tzppfl+\vorgfl$ in the temperature
equations; this is required to make the dissipation positive definite,
following the same considerations as those concerning energy
conservation resulting from the same combination under the $\dpl$ and
$\kappacv$ operators.

\subsection{Nonlinear Dissipation and the Numerical Scheme in GEM
\label{sec:cascade}}

One final dissipation mechanism remains to be considered, and in
gradient driven turbulence it is often the most important one: nonlinear
cascading to arbitrarily small scales \cite{focus}.  This enters
explicitly as an artificial diffusion term in each equation for
computations using a dissipation free scheme to calculate the nonlinear
advection terms.  

The energy cascade in drift wave turbulence is generally local in
$\kperp$-space \cite{camargo}, and can proceed in either direction
following the properties of the various nonlinearities
\cite{gang,sorgdw,camargo}.  When energy cascades to the scale of the
computational grid (highest $\kpp$ values), it must be removed somehow
lest the spectra approach the unphysical forms representing the maximum
entropy state \cite{gang} of the discrete system.  One must check to
ensure that the grid dissipation rate is independent of the resolution,
essentially the same statement contained in high Reynolds number
turbulence (dissipation independent of the diffusion or viscosity
coefficient).

In the past, the predecessor of this model has used an
upwind scheme (a slope limiting 
algorithm \cite{vanleer} integrating all the dimensions together
\cite{colella}, from computational fluid dynamics) implemented as
discussed elsewhere \cite{gyroem}.  
Herein, we employ an alternative finite difference scheme which also
does not
involve Fourier transforming or spectral operations and hence is
applicable to situations forbidding such operations.  The scheme has
been used with success by Naulin on the fluid drift Alfv\'en model
\cite{tyr}.  The first derivatives involved in Poisson bracket
structures are evaluated with the second-order version of the
Arakawa spatial discretisation
\cite{arakawa}.  The linear terms involving parallel dynamics ($\pps{}$)
and perpendicular compressibility ($\kappacv$) are evaluated with
standard second-order central differences.  Direct dissipation terms
(e.g., collision-based frictional damping of $\Jfl$ or Landau-based
damping of heat fluxes) are evaluated directly.  The entire right side
is evaluated thereby once per time step, but using a third-order
``stiffly stable'' algorithm derived by Karniadakis \etal,
according to which the previous three time
steps of the dependent variables and the right hand sides are used to
get the new time step \cite{karniadakis}.  Since the entire right hand
side is used this way, an unsplit second-order accuracy is achieved.
Finally, the artificial dissipation terms are applied separately, using
the dependent variables at the now-previous timestep.  The structure of
the equations is given by
\begin{equation}
\ptt{F} = S + D(f)
\end{equation}
where $F$ is the functional of the dependent variables $f$ appearing
under the ($\ppt{}$) operator in each equation, $S$ is the
right hand side of each equation, and $D$ is the artificial dissipation
operator in each equation.  The structure of the algorithm is given by
\begin{eqnarray}
S_0 = S(f_0) \\
F_1 = {6\over 11}\LB 3 F_0-{3\over 2} F_{-1}+{1\over 3} F_{-2}
	+\Delta t\LP 3 S_0-3 S_{-1}+S_{-2}\RP\RB \\
F_1 \from F_1 + \Delta t D(f_0) \\
\hbox{(apply boundary conditions)} \\
f_1 \from F_1
\end{eqnarray}
where the subscript `0' refers to the current timestep, `1' refers to
the new timestep, and the negative ones refer to the previous timesteps,
$\Delta t$ is the timestep interval, ``boundary conditions'' refers to
the loading of the guard cells at the computational boundary so that
derivatives are computed normally during the evaluations of $S$ and $D$,
and the last step recovering $f_1$ from $F_1$ refers to the solving of
the polarisation equations to recover $\phifl$ and
$\Afl$ and the evaluation of the gyroaveraging operators $\Gamma_1$
and $\Gamma_2$ to get the gyroreduced potentials $\phigfl$ and
$\vorgfl$.  For waves, this scheme is stable without the use of $D$,
allaying the principal consideration which led to the upwind scheme in
the first place \cite{gyroem}.  But for turbulence we require the use of
$D$.

It is important to note that the artificial dissipation must work in all
three coordinates, not just the two perpendicular ones.  
ExB advection is the main agent causing the direct cascade towards large
wavenumbers in the gyrofluid state variables, mostly $\nzfl$ (for edge
turbulence) but also $\tzplfl$ and $\tzppfl$ (almost solely, for core
turbulence).  It is important to note that this occurs not only in
$\kperp$-space but also $\kpl$-space, simply due to the statistics
\cite{albert}.  The dissipation operators must therefore function for
both $\kkpp$ and $\kkpl$.

One might be tempted to apply $D$ to the force potentials, e.g.,
$\nzfl-\phigfl$ instead of $\nzfl$, but this has been found to damage
the solution measurably.  It is indeed important not to apply artificial
dissipation directly to either $\phi$ or $\Afl$, the main effect of that
being to destroy the Alfv\'en dynamics (for $\kkpl$) or medium to large
scale vorticity (for $\kkpp$).
This was the problem
with the upwind scheme \cite{gyroem}: as the kinetic shear Alfv\'en
velocity is scale dependent the exact one could not be used in the flux
splitting involved in the scheme, so the fastest one was used (otherwise,
the scheme is unstable).
For $\beta_e<m_e/M_i$ the fastest wave (following $v_A$)
is at the lowest $\kpp$ and the
smallest scales (highest $\kpp$) are dominated by collisional
dissipation anyway (since $\nu_e>c_s/\Lpp$), so edge turbulence was not
strongly impacted.  For core turbulence, on the other hand, the fastest
wave (following $V_e$) is at the highest $\kpp$ so that the large scale
MHD response at the lowest $\kpp$ is strongly dissipated with a sort of
super-resistivity acting directly upon $\Afl$ rather than $\Jfl$.  To
avoid the same problem with schemes with explicitly applied artificial
dissipation, it is important to avoid
application of any of the artificial dissipation operators directly to
$\Afl$.

In the
$xy$-plane the operations are summarised by the statement
\begin{equation}
\uedl \to \uedl - \nupp\ddpp - \nupl\ddpl
\end{equation}
in each equation; that is, artificial dissipation in both the $xy$-plane
and the $s$-direction is
applied to whatever is advected by the gyroreduced ExB velocity.
An alternative is a hyperdiffusion for the $xy$-plane, so that
\begin{equation}
\uedl \to \uedl + \ddpp\nupp\ddpp - \nupl\ddpl
\end{equation}
is used.  For models with variable $B$ these should be 
respectively changed to
\begin{equation}
\uedl \to \uedl - \div\nupp\rs^2\dpp - \div(\bunit\nupl\bunit)\cdot\grad
\end{equation}
and
\begin{equation}
\uedl \to \uedl + \ddpp\nupp\rs^4\ddpp - \div(\bunit\nupl\bunit)\cdot\grad
\end{equation}
with
\begin{equation}
\rs^2 = {1\over B^2}
\end{equation}
in normalised units,
so that the property of positive definiteness is preserved.  If variable
resolution causes problems, then the metric elements in these forms
should be replaced by their flux surface averages.

Typical values of these dissipation coefficients are set depending on
the physical situation; in general they must be set as small as
possible.  Full resolution is found when it can be shown the resulting
grid dissipation rate (not necessarily the answer for the transport
fluxes) is independent of the dissipation parameters.  A resolution
study will generally not be done at a particular value of the
coefficient; rather, the coefficient should be made smaller when the
resolution is increased.  A window of operation opens when it is
subsequently found that the above criteria for full resolution is met.
Tests on core turbulence with adiabatic electrons
($\nu_e=\beta=\mu_e=0$) find that $\nupp$ as
small as $10^{-2}$ is possible with resolutions of $h_x=h_y=1$ or
$2\times\rs$.  Edge turbulence 
($C,\bhat,\muhat$ all unity or greater; cf.\ Section \ref{sec:norm} and
Ref.\ \cite{eps03}) 
requires $h_x=h_y=1\times\rs$ or smaller
to be able to reduce $\nupp$ to as small as $3\times 10^{-2}$.  With
$h_x=h_y=2\times\rs$ a value of $\nupp=0.1$ can be required, and this is
generally too large to allow the vorticity dynamics in the range
$0.5<\kpp\rs<1$ to function properly.  This is due to the robust
nonlinear action by $\vedl\nefl$ in edge turbulence \cite{focus}.  Under
these conditions the hyperdiffusion form is necessary to be able to
reproduce the nonlinear drift wave instability.  For cold ion models
($\tau_i=0$) with no temperature dynamics, this instability can be
reproduced with a resolution of $h_x=h_y=2$ and a hyperdiffusion of
$\nupp=0.01$. 

The
parallel 
dissipation coefficients are easier as they are only needed to contain
the cascade in the parallel wavenumber $\kpl$ by the nonlinear
perpendicular dynamics.  Values of $\nupl=3\times 10^{-3}$ for both ions and
electrons are found to be sufficient with $h_s=2\pi qR/16$, and for the
most important wavenumber range $-2<\kpl qR<2$ these lead to small
corrections to the physical dissipation rates.  It has been found
necessary to use the same coefficient for both ions and electrons, to
avoid artificial charge separation which can have a large effect on the
spectral region with $\kpl qR$ moderate and $\kpp\rs$ small.

\section{Selected Computational Results
\label{sec:results}}

It is not the purpose of this paper to enter detailed study of any of
the problems the GEM model is to be applied to; rather, the focus is
upon the way energetics works in the model and to use that to assist
consistent construction of the model.  Nevertheless, it is useful to
apply the model briefly herein to an elementary situation whose capture
is important (kinetic shear Alfv\'en wave damping \cite{wwleealfven,tilman}), a
well known set of computational results (the Cyclone ITG turbulence
campaign \cite{dimits}), and a demonstration that electron driven
turbulence at scales below the ion gyroradius can be addressed with a
gyrofluid model (``ETG'' \cite{jenkoetg}).  The latter two cases will be
treated in proper detail in the future.  ETG turbulence has been treated
with a fluid model before, but only with adiabatic ion models
\cite{hortonetg,drakeetg,ottavianietg}.  Herein, we apply GEM directly
and find the ETG dynamics occurring naturally at its native scales.

\subsection{Kinetic shear Alfv\'en wave damping}

Shear Alfv\'en waves are well known from MHD \cite{freidberg}, and the
collisionless kinetic counterpart (KALF)
is also well known \cite{kalf}.  It
has already been shown the gyrokinetic model treats them properly
\cite{wwleealfven,tilman}.  We now use the result to calibrate the model Landau
damping coefficient for the electrons.

We take a basic parameter case with $\bhat=1$ and $\muhat=1$ and
$\epss=18350$ with both collisionalities set to zero as a reference.
The magnetic field is straight and homogeneous ($g^{xx}=g^{yy}=B=1$ with
$\shat=0$ and $\kappacv=0$) and there are no background gradients
($\wn=\wt=\wi=0$).  The ions are cold ($\tau_i=0$ hence $\rho_i=0$).
The Pad\'e version of the local model is used.  With the homogeneous
situation, the profile functions are set to zero and the domain is
periodic in both $y$ and $s$.
The perpendicular 
domain sizes are $L_x=L_y=2\pi/K$, with $K=0.1$.
The parallel domain is one connection length.  
The initial state is the sinusoidal disturbance 
$\nefl=10^{-4}\LP 1+\cos Kx\RP\cos Ky\cos s$,
with $\nifl=\nefl$.
The grid was $32\times 32\times 16$ in $\{x,y,s\}$.  
The values of $\bhat$ and $\muhat$, and the Landau damping model
coefficient $\ald$, were varied between $0.1$ and $10$.
Artificial dissipation ($\nupp$ and $\nupl$) was set to zero.

\begin{figure}[H]
\centerline{\epsfxsize=5 true cm\epsfbox{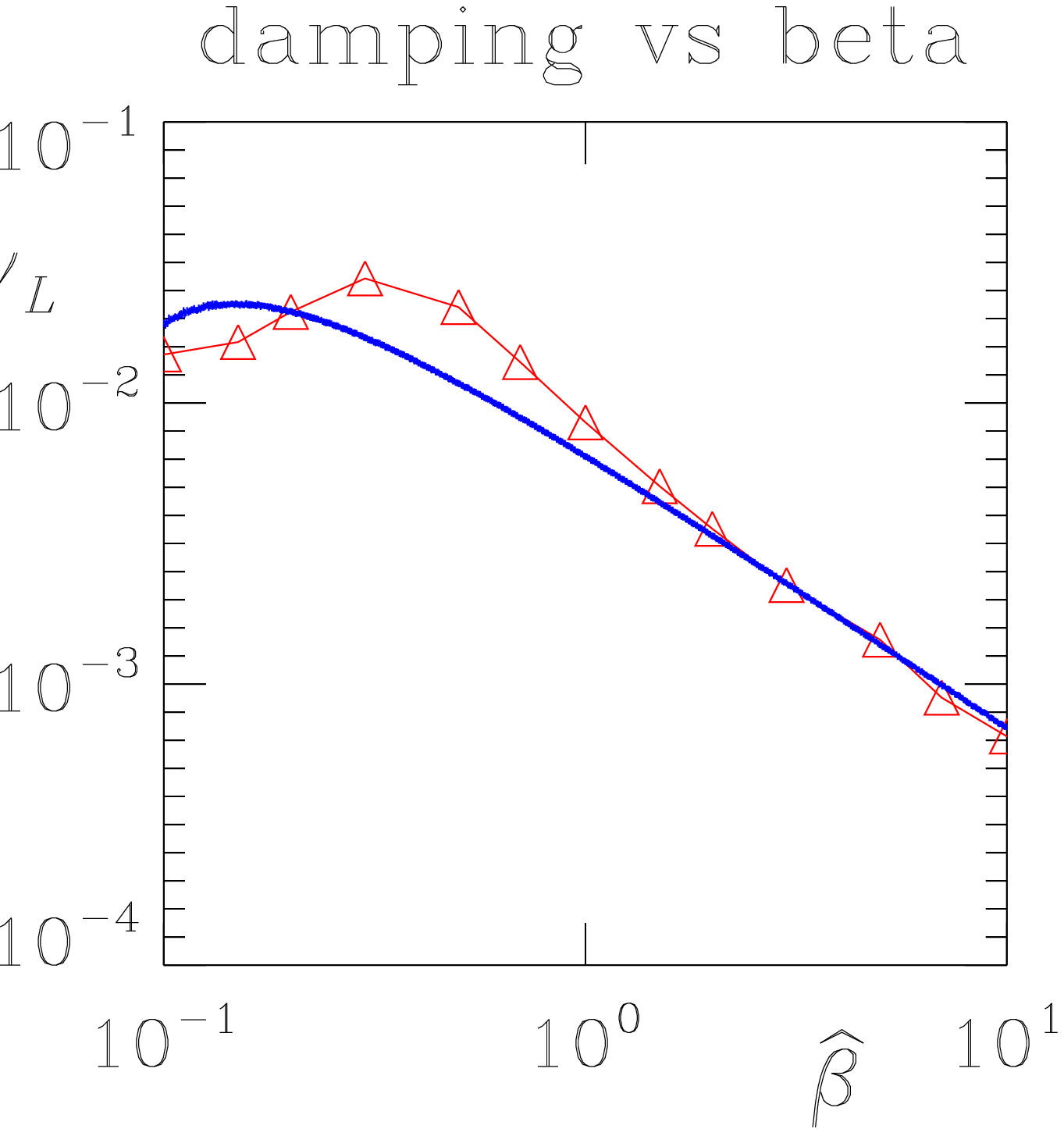}\hskip 0.5 cm
	\epsfxsize=5 true cm\epsfbox{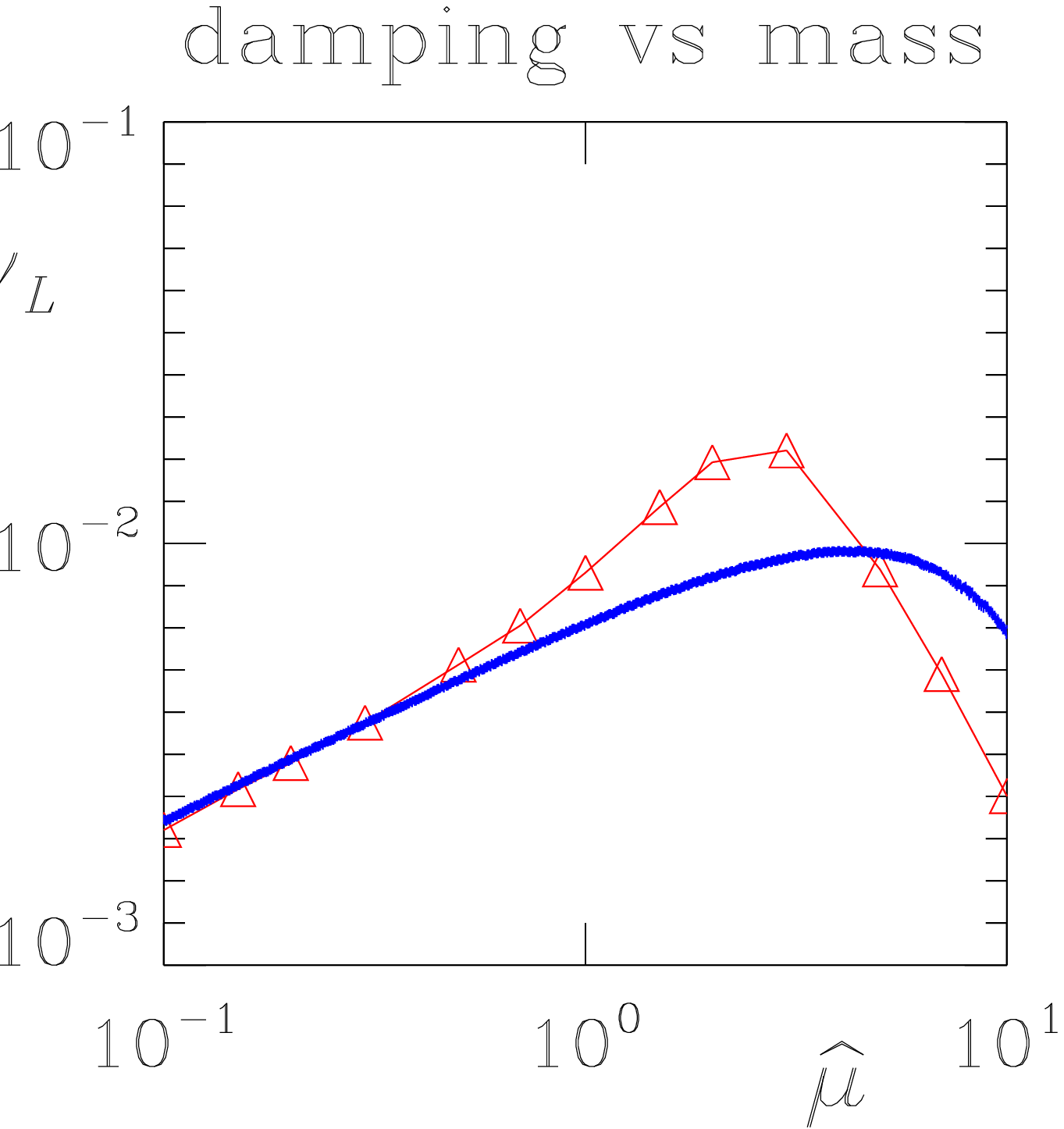}\hskip 0.5 cm
	\epsfxsize=5 true cm\epsfbox{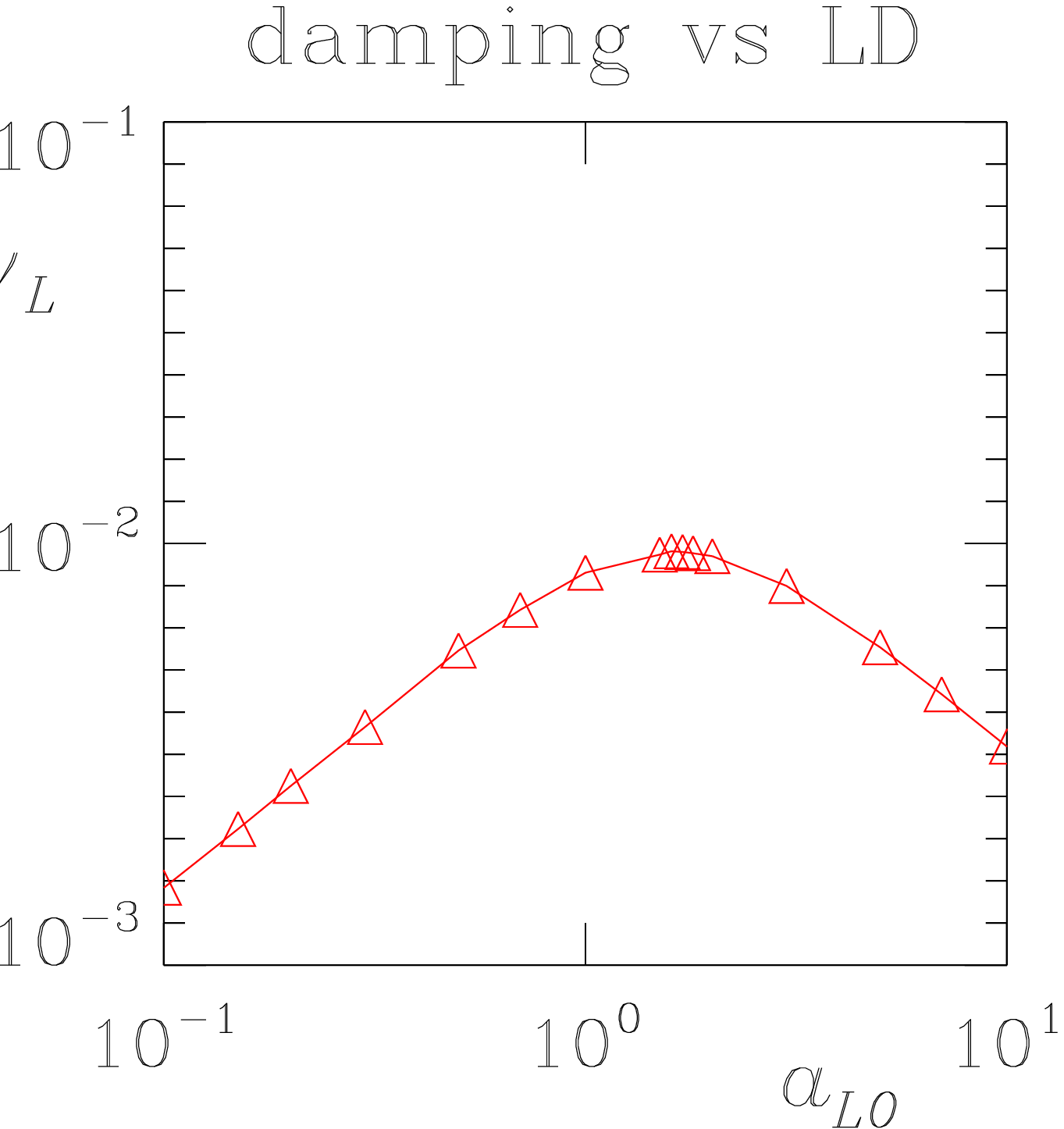}}
\null\vskip 1 cm
\caption{
Kinetic shear Alfv\'en damping rates versus normalised beta, electron
mass, and Landau damping closure coefficient.
In the leftmost two frames the blue line (whose peak is toward lower
$\bhat$ values and higher $\muhat$ values)
gives the kinetic dispersion relation using the method of Ref.\ \cite{tilman}.
The calibration works in the $\beta_e\gg\mu_e$ regime and
yields qualitatively similar behaviour elsewhere, though the details of
the peaks can only be captured with a kinetic model.
}
\label{figkalf}
\end{figure}

The KALF dispersion relation, shown in Fig.\ \ref{figkalf},
has the two standard asymptotic limits
$\bhat/\muhat\gg 1$ and $\kkpp\ll 1$ where for these cases $\kkpp=(5/4)K^2$.
This range is found with the nominal sweep in $\bhat$ for $\bhat\gg 1$,
and in the sweep in $\muhat$ for $\muhat\ll 1$, and is well captured by
the GEM model as shown by the comparison to the kinetic result using the
root finding method of Ref.\ \cite{tilman}.
When $\bhat\approx\muhat$ there is substantial thermal electron
resonance.  In this regime the GEM model shows a peak, but with the peak
value and its location only approximately captured.
In the sweep of $\ald$ the damping rate was proportional to
$\ald$ only for $\ald<1$.  For $\ald>1$ the effect is to remove the
parallel heat flux from the dynamics, which becomes more ideal; hence
the damping rate falls again.  The maximum is found for $\ald=1.6$,
which is close to the actual thermal resonance at $\sqrt{3}$.  The
damping rate varies within the interval 
$8.5<-10^3\gamma_L<9.6$
for $1<\ald<2$.
As a robust model in the absence of fitting for all possible cases, it
appears to be sufficient to simply leave $\ald=1$, and the model
performs qualitatively well.

\subsection{ITG turbulence}

The standard of core turbulence studies with adiabatic electrons is the
Cyclone project, which benchmarked a series of models and computations
against a particular case of hot ion collisionless turbulence and
transport \cite{dimits}.  With the only free energy source being the ion
temperature gradient, this is called ITG turbulence.  With adiabatic
electrons taken as a model ($\bhat=\muhat=C=0$), the parameter set is
given by
\begin{eqnarray}
  \wcv=0.290 \qquad
  \epss=93.4 \qquad
  \shat=0.78 \nonumber\\
  \wt=\wi=\tau_i=1 \qquad
  \wn=0.321
\label{eqitgparms}
\end{eqnarray}
The artificial dissipation coefficients are $\nupp=0.01$, using
hyperdiffusion, 
and $\nupl=0.001$.  The boundary dissipation coefficient was $1.0$.
For the magnetic field 
the simple circular tokamak model with globally consistent boundary
conditions and shifted metric coordinate system
is used as detailed in Ref.\ \cite{shifted}.
The potential is initialised with
a random disturbance bath in $x$ and $y$ \cite{ssdw} and a parallel
envelope following the field lines from $s=0$ \cite{shifted}, and an RMS
amplitude of $10^{-8}$.
The Pad\'e version of the global model is used.
The basic profile is given by  
\begin{equation}
p_0(x) = {L_x\over 2}\LP 1 - \sin {\pi x\over L_x}\RP
\end{equation}
and then both densities are initialised with $\wn p_0+\phifl$ and the
ion temperatures ($\tiplfl,\tippfl$) with $\wi\tau_i p_0$.  The
adiabatic form of the polarisation equation with profiles is used, with
$\avg{\nefl}=\wn p_0(x)$, noting that $\wt$ has no role for
this problem.
The perpendicular 
domain sizes are $L_x=L_y=80\pi$, roughly commensurate with the global
tokamak 
dimensions of $a/\rs=192$ and $2\pi r/q\rs = 350$ (where $r=a/2$).  The
parallel domain is one connection length.  The grid was $128\times
128\times 16$ in $\{x,y,s\}$.

The four cases with $\wi=\{0.8,1.0,1.5,2.2\}$ were taken.  Normalisation
of the transport level is to $L_n$, following Ref.\ \cite{dimits}, so
that the transport flux in units of the nominal $\Lpp$,
$Q_i=\avg{(0.5\tiplfl+\tippfl)\vex}$, is recast in terms of a transport
coefficient by taking $\chi_i = Q_i/(\wn\abs{\grad T})$.  Each run begins
in a linear growth phase, overshoots to a transport level in the
vicinity of $\chi_i=10$, and then saturates with the zonal flow dynamics
(the part of the ExB flow arising from $\avg{\phifl}$) reaching
statistical equilibrium only well after $t=1000$.  Runs were taken to
$t=4000$.  The transport is displayed statistically, with a sample taken
at intervals of $\Delta t=10$ in the phase $1000<t<4000$.  Slow
relaxation of the temperature profile fills out the transport scaling
curve.  For each sample, the flux and gradient were averaged over all
grid nodes in the part of the domain with $0<x<L_x/4$ before their
ratio was computed.

\begin{figure}[H]
\centerline{\epsfxsize=10 true cm\epsfbox{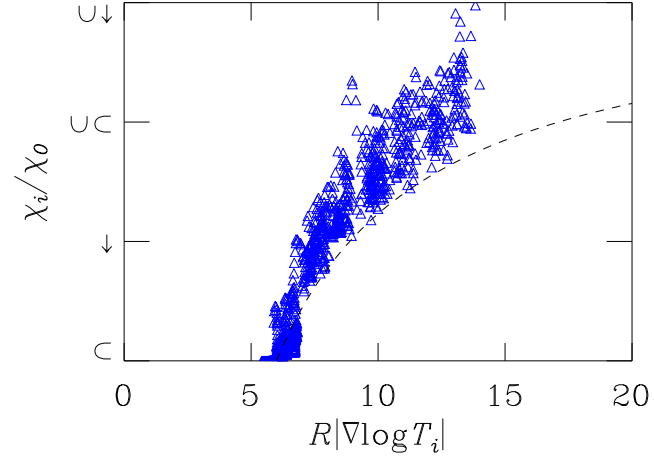}}
\null\vskip 1 cm
\caption{
Transport of ITG turbulence found by GEM (triangles, one per sample as
described in the text), compared to the gyrokinetic fit from Ref.\
\cite{dimits}.  The transport diffusivity, $\chi_i$, calculated
temporally as described in the text, is normalised to a nominal value of
$\chi_0=\rs^2 c_s/L_n$.  The fact that the
groups of points from four decaying runs overlap well indicates the
transport to be temporally local.
}
\label{figchii}
\end{figure}

The resulting transport curve is shown in Fig.\ \ref{figchii}, wherein
the triangle markers denote each sample (1200 in all), and the dashed
curve is the fit to the gyrokinetic particle model results as given in
Ref.\ \cite{dimits}.  Agreement at the $20\%$ level is found for most of
the curve, and moreover the nonlinear threshold agrees within the
statistical scatter.  Moreover, the fact that the groups of points from
four decaying runs overlap well indicates the transport to be temporally
local.

\subsection{ETG turbulence}

A class of turbulent dynamics at scales smaller than $\rho_i$ driven by
$\grad T_e$ is called ETG \cite{jenkoetg}.  Neither $\grad n$ nor $\grad
T_i$ is available as a drive because the ions are adiabatic (in the
simplest treatments) or nearly so.  Here, we carry both electrons and
ions with the full six moments and allow the spatial scale range kept in
the particular case to determine the dynamics.  
The Fourier version of the local model is used,
with profile functions $n_e=n_i=-\wn x$ and $T_e=-\wt x$
and $T_i=-\wi x$.  
The same magnetic geometry as in the ITG examples above is used.  The
same random bath as above is used initially, but for $\nefl$.

\begin{figure}[H]
\centerline{\epsfxsize=8 true cm\epsfbox{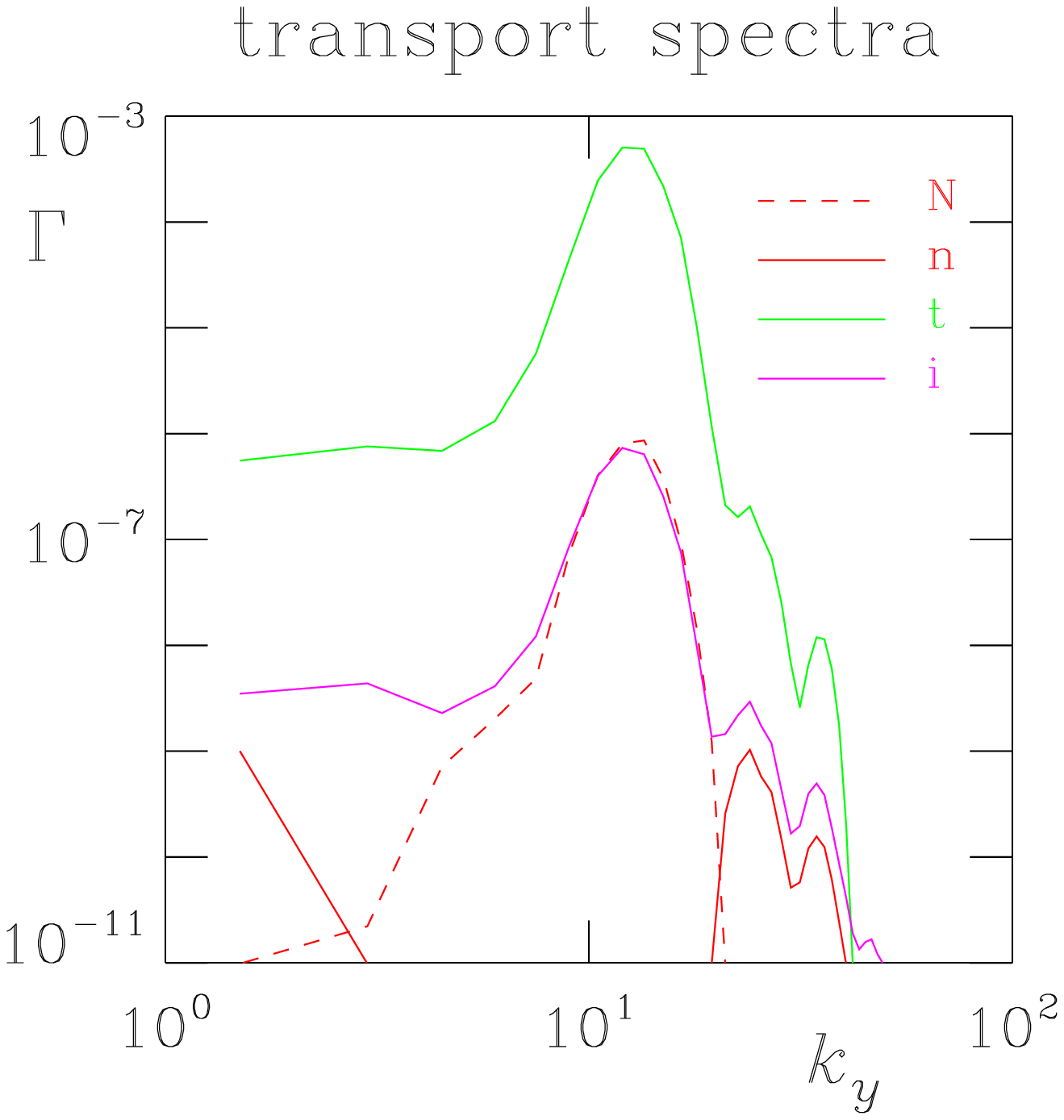}\hskip 0.5 cm
	\epsfxsize=8 true cm\epsfbox{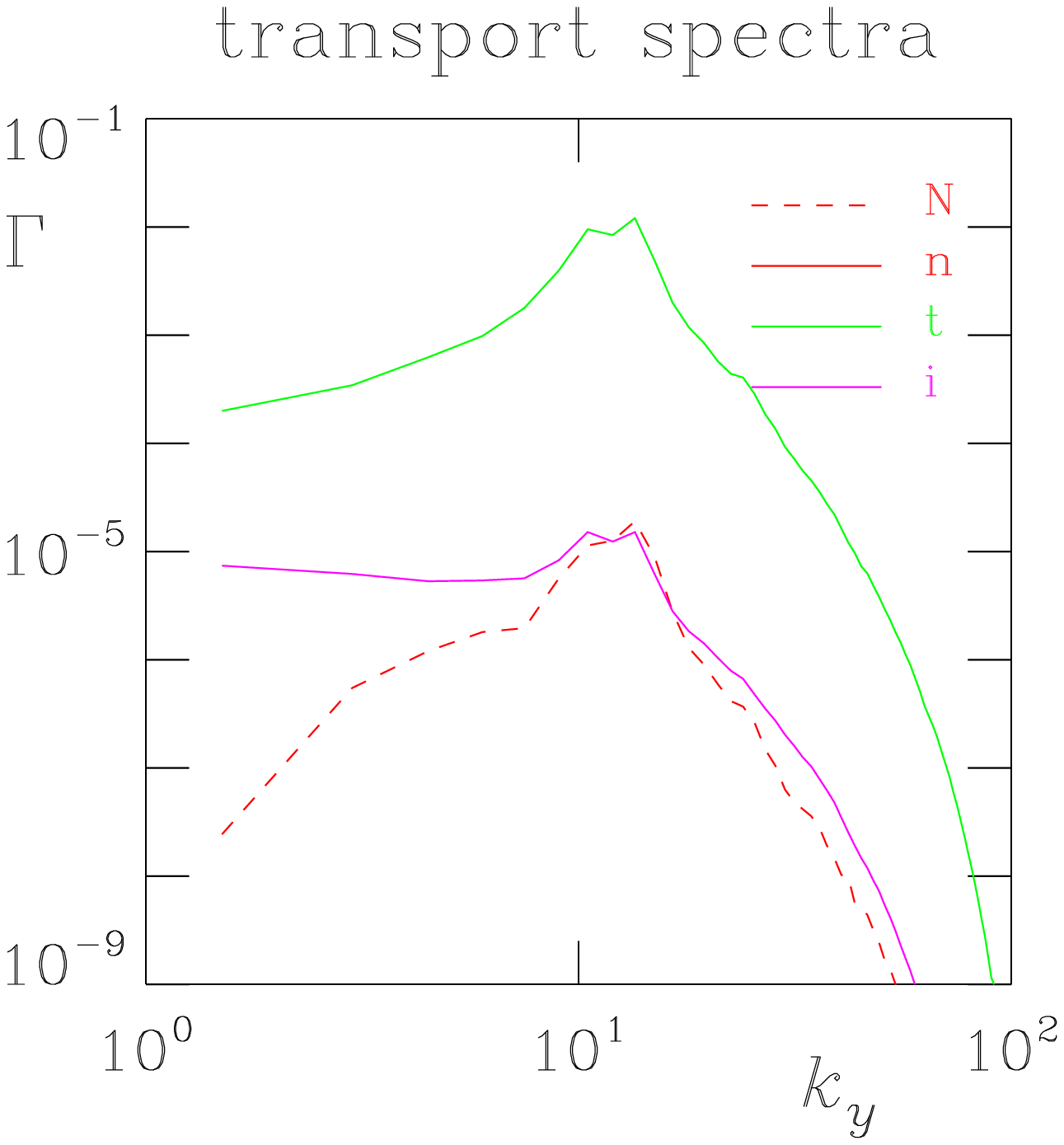}}
\centerline{\epsfxsize=8 true cm\epsfbox{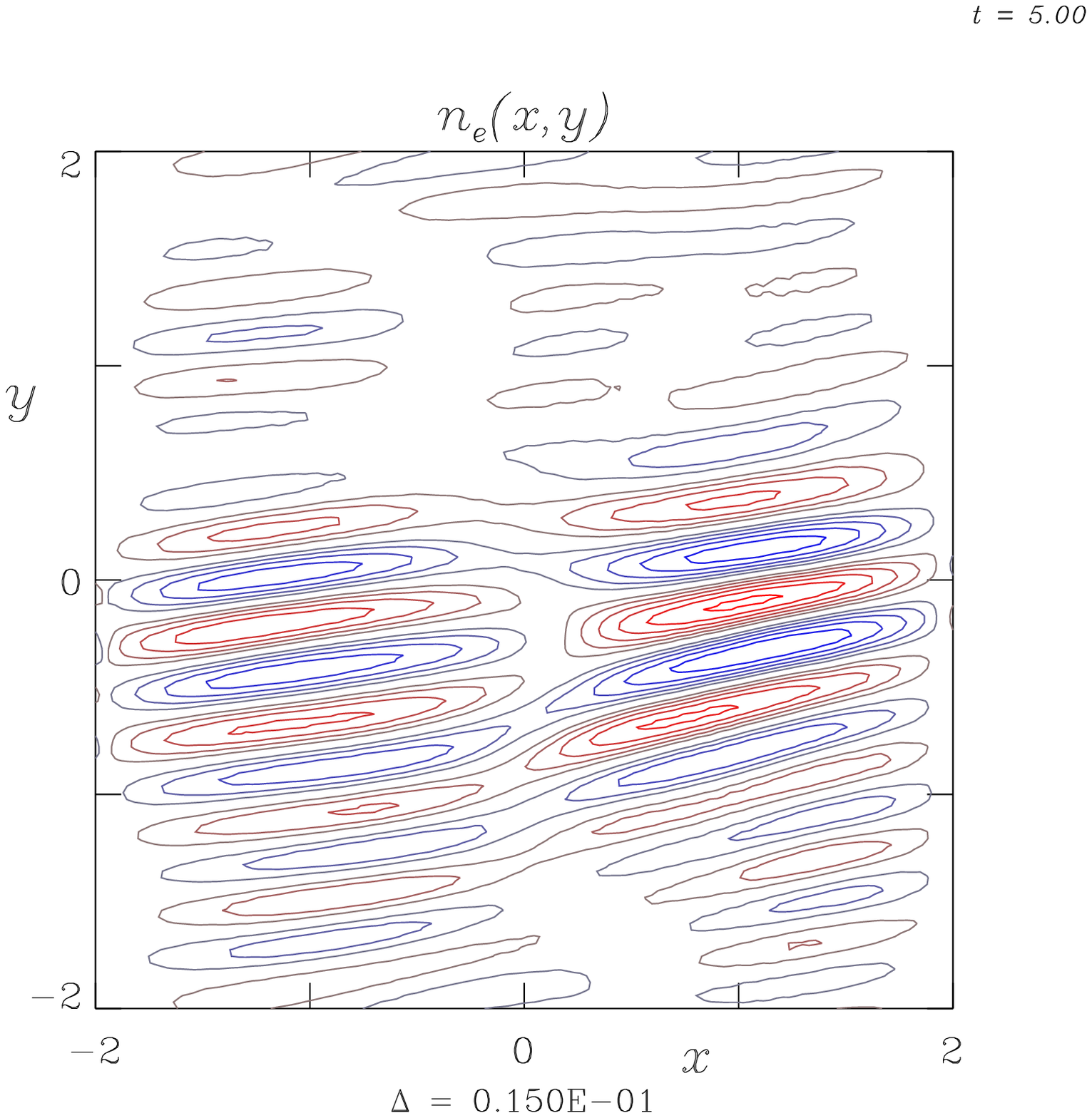}\hskip 0.5 cm
	\epsfxsize=8 true cm\epsfbox{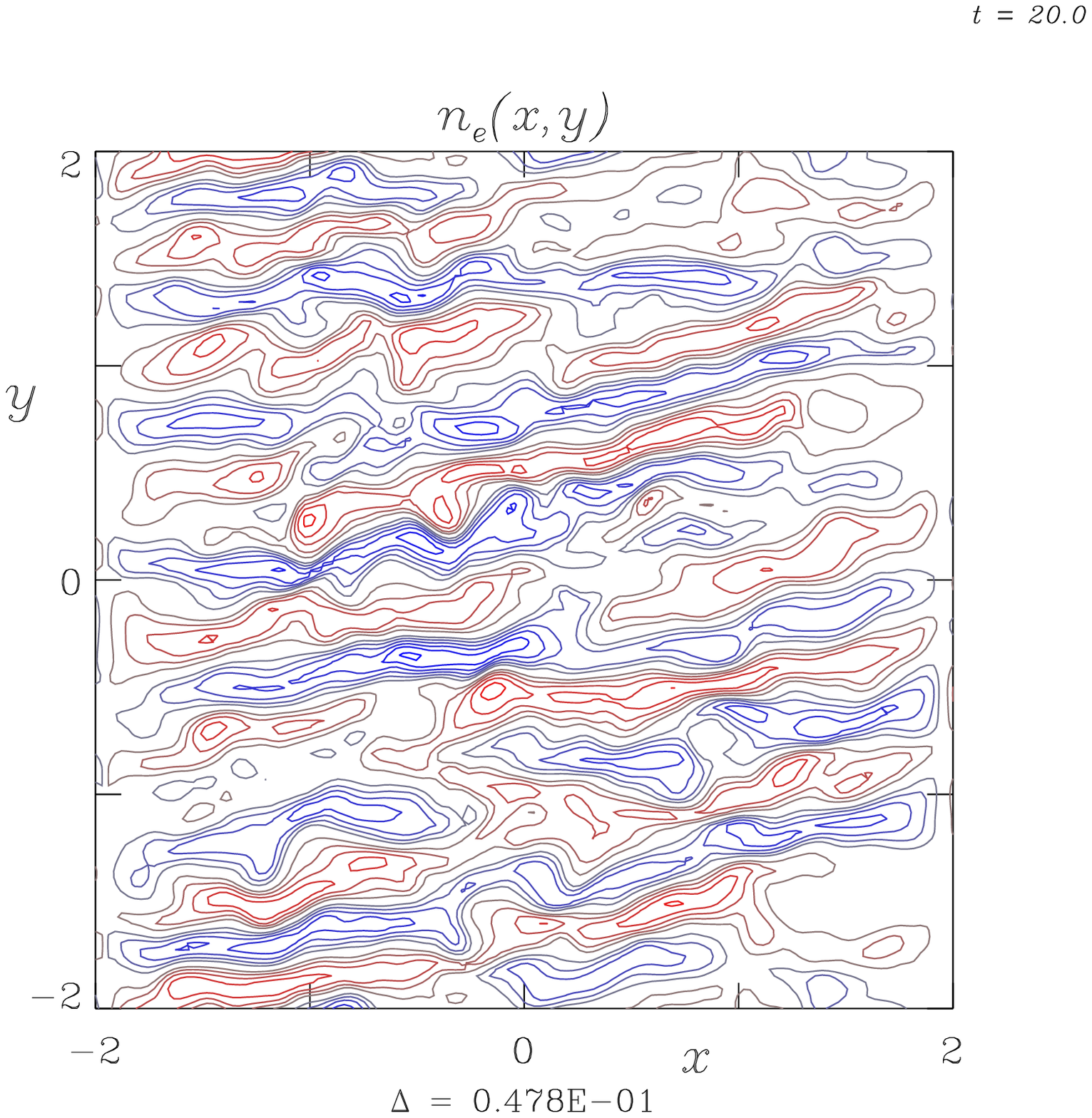}}
\null\vskip 1 cm
\caption{
Transport spectra and density morphology in core ETG turbulence in the
linear (left) and saturated (right) phases, as described in the text.
Lines marked 'n' and 'N' are for the particle flux, where it is positive
or negative, respectively.  Lines marked 't' and 'i' are for the
electron and ion conductive heat fluxes, respectively.
The scales are normalised to $\rs$; multiply $\{x,y\}$ and divide $k_y$
by $60.6$ to obtain them in terms of $\rho_e$.
}
\label{figetg}
\end{figure}

Here we merely demonstrate the ability of the GEM model to capture this
ETG turbulence for typical core parameters,
the same as the one used for the ITG examples above, additionally with
$\bhat=0.464$ and $\muhat=0.0254$ and $\nu_e=0.0333$.  The artificial
dissipation coefficients were $\nupp=3\times 10^{-3}$ (using simple
diffusion, not hyperdiffusion) and
$\nupl=10^{-4}$.  The initial RMS amplitude for $\nefl$ was $a_0=3\times
10^{-3}$.  The spatial domain size was $L_x=L_y=4\pi/3$ for the drift
plane and one connection length along the magnetic field.  The grid was
$128\times 128\times 16$ in $\{x,y,s\}$.  The timestep was $5\times
10^{-4}$.  The run was carried for $20 \Lpp/c_s$.  With the minimum
value of $k_y\rho_i$ of $1.5$, ITG activity is generally absent.  The
fastest growing spectral range is about $10<k_y<20$, representing
structure 
scales $\Delta y=\pi/k_y$ on the order of $10\rho_e$.  The spatial
morphology shows these to be
radially extended, with $\Delta x>8\Delta y$.  The nonlinear transition
begins at $t\approx 8$ and saturation occurs after $t\approx 12$.
A very strong transport level is found, just under $0.1$, which in terms
of electron scales is $\chi_e\approx 6\rho_e^2V_e/L_T$.  The spectrum is
broader than in the linear phase, but still narrow compared to edge
turbulence, and more importantly the transport spectrum peaks at
$k_y=10$, very close to the linear growth peak.  The radially extended
structures do persist in the saturated phase, by contrast to typical
core ITG or edge turbulence.  These features, shown for both linear and
nonlinear phases in Fig.\ \ref{figetg}, are the same as those shown
previously by nonlinear gyrokinetic studies \cite{jenkoetg}.  The
amplitude of the electron moment variables $\nefl$ and $\teplfl$ is
about $0.2$, while the corresponding ion variables are about two orders
of magnitude smaller.

\section{Summary}

The standard local gyrofluid model has been placed on energetically
consistent grounds, with the moment variables and the electrostatic
potential given a full finite Larmor radius (FLR) treatment at the same
level of sophistication.  The FLR effects on the magnetic potential and
parallel velocities and heat fluxes is left to the future.  With these
changes it is possible to recover important results emerging from
gyrokinetic computations, with a computationally more tractable model.
Large systems may be treated with full resolution with what at present
time may be regarded as modest computational resources.  The model is
flexible, to the extent that the level of sophistication can be
increased or decreased while retaining energetic and geometric
consistency.  Both global and local situations can be treated.  Highly
detailed dissipative linear closures as discussed in the main references
\cite{dorland,beer} are not necessary in many cases of interest, in
particular the one from the Cyclone study (Ref.\ \cite{dimits}).

For proper edge turbulence ($\muhat>1$ and $C>1$) in the electromagnetic
regime ($\bhat>1$) the model functions much as in previous versions as
published elsewhere \cite{gyroem}.  Results from the two moment version
GEM3 (density and parallel velocity for both electrons and ions) are
published elsewhere \cite{eps03}, showing the role of the three
dimensional drift wave nonlinear instability in the context of tokamak
edge turbulence as done previously \cite{focus} for the corresponding
fluid model.  Work with cases with various ratios of $\eta_i=\wi/\wn$
(cf.\ \cite{focus} for the role of this in the fluid model) and with two
ion species is in progress.

\par\vfill\eject

\appendix

\section{Simple Correspondence between Fluid and Gyrofluid Models
\label{sec:corresp}}

A simple exercise using the most basic reduced MHD interchange model
helps gain insight into the relationship between the fluid and gyrofluid
models.  We start with the two equations written down in the
Introduction, writing the electron density equation in terms of a charge
density to equalise the units.  For reasons which will become clear, we
retain the diamagnetic compression effect in the density equation.  We
also incorporate the profile variation of the thermal state variables,
normally acted up solely by ExB advection or magnetic flutter, into the
corresponding dependent variables (the only difference this makes is
that $\kappacv$ now acts upon the profiles, which is actually somewhat
more realistic).  The equations are
\begin{eqnarray}
{n_iM_i c^2\over B^2}\dtt{}\ddpp\phifl = - T_e\kappacv(\nefl) \\
e\dtt{\nefl} = n_e e\kappacv(\phifl) - T_e\kappacv(\nefl)
\end{eqnarray}
with $d/dt$ representing the ExB advective derivative.
We now define arbitrarily an auxiliary variable, $\Nfl$, as
\begin{equation}
\Nfl e = \nefl e - {n_iM_i c^2\over B^2}\ddpp\phifl
\end{equation}
The evolution equation for $\Nfl$ is found therefore by subtracting 
Eqs.\ (A1,A2),
\begin{equation}
e\dtt{\Nfl} = n_e e\kappacv(\phifl) 
\end{equation}
By inspection with any of the gyrofluid models, we find that $N$ is
simply the gyrocenter ion density $n_i$, with the sole proviso that the
background constant parameters for $n_e$ and $n_i$ are equal.
This tells us that the MHD
formulation for the ExB 
vorticity is identical to the cold-ion limit of the polarisation density
in the gyrofluid model.  The relation here is between the polarisation
current in the fluid model and the polarisation density in the gyrofluid
model.
The MHD interchange term $\kappacv(\nefl)$
appears properly only if the diamagnetic compression effect is kept in
the gyrofluid density equations (i.e., unlike for a fluid model, this
effect cannot be neglected in a gyrofluid model).  The ExB compression
effect in the density cancels out of the interchange effect in the
vorticity.  However, it is necessary in either model to retain the ExB
compression in order to conserve energy.  The incidental benefit is to
retain the geodesic curvature effect which is the principal mechanism
limiting the growth of zonal flows \cite{gdcurv}.

More detailed accounts of the correspondence between the fluid and
gyrofluid models may be found elsewhere \cite{dorland,eps03,belova}.

\par\vfill\eject

\section*{Figure Captions}

{\tt Figure 1.}\emskip
Kinetic shear Alfv\'en damping rates versus normalised beta, electron
mass, and Landau damping closure coefficient.
In the leftmost two frames the blue line (whose peak is toward lower
$\bhat$ values and higher $\muhat$ values)
gives the kinetic dispersion relation using the method of Ref.\ \cite{tilman}.
The calibration works in the $\beta_e\gg\mu_e$ regime and
yields qualitatively similar behaviour elsewhere, though the details of
the peaks can only be captured with a kinetic model.

{\tt Figure 2.}\emskip
Transport of ITG turbulence found by GEM (triangles, one per sample as
described in the text), compared to the gyrokinetic fit from Ref.\
\cite{dimits}.  The transport diffusivity, $\chi_i$, calculated
temporally as described in the text, is normalised to a nominal value of
$\chi_0=\rs^2 c_s/L_n$.  The fact that the
groups of points from four decaying runs overlap well indicates the
transport to be temporally local.

{\tt Figure 3.}\emskip
Transport spectra and density morphology in core ETG turbulence in the
linear (left) and saturated (right) phases, as described in the text.
Lines marked 'n' and 'N' are for the particle flux, where it is positive
or negative, respectively.  Lines marked 't' and 'i' are for the
electron and ion conductive heat fluxes, respectively.
The scales are normalised to $\rs$; multiply $\{x,y\}$ and divide $k_y$
by $60.6$ to obtain them in terms of $\rho_e$.

\par\vfill\eject

\par\vfill\eject

\par\vfill\eject

\end{document}